\documentclass[a4paper,11pt]{article}

\usepackage{jinstpub} 

\usepackage{subcaption}

\title{SYNCA: A Synthetic Cyclotron Antenna for the Project 8 Collaboration}

\newcommand{\mitt}{Laboratory for Nuclear Science, Massachusetts Institute of Technology, Cambridge, MA 02139, USA}
\newcommand{\mainz}{Institute for Physics, Johannes Gutenberg University Mainz, 55128 Mainz, Germany}
\newcommand{\psu}{Department of Physics, Pennsylvania State University, University Park, PA 16802, USA}
\newcommand{\cwru}{Department of Physics, Case Western Reserve University, Cleveland, OH 44106, USA}
\newcommand{\uw}{Center for Experimental Nuclear Physics and Astrophysics and Department of Physics, University of Washington, Seattle, WA 98195, USA}
\newcommand{\pnnl}{Pacific Northwest National Laboratory, Richland, WA 99354, USA}
\newcommand{\yale}{Wright Laboratory, Department of Physics, Yale University, New Haven, CT 06520, USA}
\newcommand{\llnl}{Lawrence Livermore National Laboratory, Livermore, CA 94550, USA}
\newcommand{\iu}{Center for Exploration of Energy and Matter and Department of Physics, Indiana University, Bloomington, IN 47405, USA}
\newcommand{\kit}{Institute for Astroparticle Physics, Karlsruhe Institute of Technology, 76021 Karlsruhe, Germany}

\author[a]{A.~Ashtari Esfahani} 
\author[b]{S.~B\"oser} 
\author[c]{N.~Buzinsky}
\author[d]{M.~C.~Carmona-Benitez}
\author[a]{C.~Claessens} 
\author[d]{L.~de~Viveiros}
\author[b]{M.~Fertl}
\author[c]{J.~A.~Formaggio}
\author[e]{L.~Gladstone}
\author[f]{M.~Grando} 
\author[a]{J.~Hartse}
\author[g]{K.~M.~Heeger} 
\author[f]{X.~Huyan} 
\author[f]{A.~M.~Jones} 
\author[h]{K.~Kazkaz}
\author[c]{M.~Li}
\author[b]{A.~Lindman}
\author[b]{C.~Matth\'{e}} 
\author[e]{R.~Mohiuddin} 
\author[e]{B.~Monreal} 
\author[d]{R.~Mueller}
\author[g]{J.~A.~Nikkel} 
\author[a]{E.~Novitski} 
\author[f]{N.~S.~Oblath} 
\author[c]{J.~I.~Pe\~{n}a}
\author[i]{W.~Pettus} 
\author[b]{R.~Reimann}
\author[a]{R.~G.~H.~Robertson} 
\author[g]{L.~Salda\~na} 
\author[g]{P.~L.~Slocum}
\author[c]{J.~Stachurska}
\author[e]{Y.-H.~Sun}
\author[g]{P.~T.~Surukuchi} 
\author[g]{A.~B.~Telles}
\author[b]{F.~Thomas} 
\author[f]{M.~Thomas}  
\author[b]{L.~A.~Thorne} 
\author[j]{T.~Th\"ummler} 
\author[h]{L.~Tvrznikova} 
\author[c]{W.~Van~De~Pontseele}
\author[f,a]{B.~A.~VanDevender} 
\author[g]{T.~E.~Weiss} 
\author[d]{T.~Wendler} 
\author[c]{E.~Zayas} 
\author[d, 1]{A.~Ziegler, \note{Corresponding author.}}

\affiliation[a]{\uw}
\affiliation[b]{\mainz}
\affiliation[c]{\mitt}
\affiliation[d]{\psu}
\affiliation[e]{\cwru}
\affiliation[f]{\pnnl}
\affiliation[g]{\yale}
\affiliation[h]{\llnl}
\affiliation[i]{\iu}
\affiliation[j]{\kit}

\emailAdd{ziegler@psu.edu}

\abstract{Cyclotron Radiation Emission Spectroscopy (CRES) is a technique for measuring the kinetic energy of charged particles through a precision measurement of the frequency of the cyclotron radiation generated by the particle's motion in a magnetic field. The Project 8 collaboration is developing a next-generation neutrino mass measurement experiment based on CRES. One approach is to use a phased antenna array, which surrounds a volume of tritium gas, to detect and measure the cyclotron radiation of the resulting $\beta$-decay electrons. To validate the feasibility of this method, Project 8 has designed a test stand to benchmark the performance of an antenna array at reconstructing signals that mimic those of genuine CRES events. To generate synthetic CRES events, a novel probe antenna has been developed, which emits radiation with characteristics similar to the cyclotron radiation produced by charged particles in magnetic fields. This paper outlines the design, construction, and characterization of this Synthetic Cyclotron Antenna (SYNCA). Furthermore, we perform a series of measurements that use the SYNCA to test the position reconstruction capabilities of the digital beamforming reconstruction technique. We find that the SYNCA produces radiation with characteristics closely matching those expected for cyclotron radiation and reproduces experimentally the phenomenology of digital beamforming simulations of true CRES signals.} 

\collaboration{Project 8  Collaboration}

\begin{document}
\maketitle
\flushbottom

\section{Introduction}
\label{sec:intro}

Neutrinos are the most abundant standard model fermions in our universe, but due to weak interaction cross-sections with other particles, neutrinos are particularly difficult to study. Consequently, many fundamental properties of neutrinos are still unknown including the absolute scale of the neutrino mass \cite{Workman:2022ynf}. Direct, kinematic measurements of the neutrino mass are particularly valuable due to their model independent nature \cite{FORMAGGIO20211}. To date the most sensitive direct neutrino mass measurements have been performed by the KATRIN collaboration \cite{Aker2022}, which measures the molecular tritium $\beta$-decay spectrum to infer the neutrino mass. Current data from neutrino oscillation measurements \cite{Workman:2022ynf} allow for neutrino masses significantly smaller than the design sensitivity of the KATRIN experiment; therefore, there is a need for new technologies for performing direct neutrino mass measurements to probe lower neutrino masses.

\begin{figure}[htbp]
\centering
\includegraphics[width=.6\textwidth]{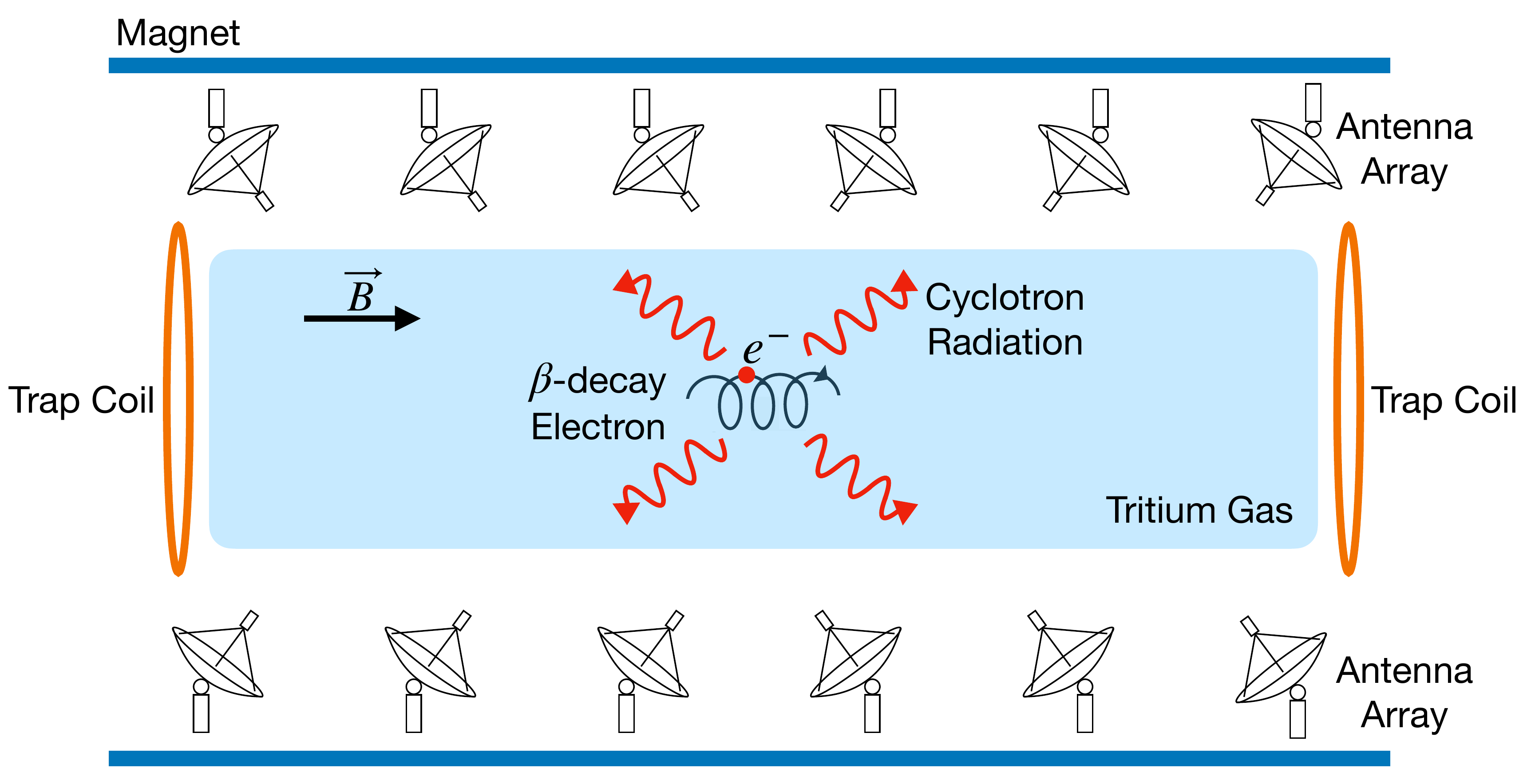}
\qquad
\caption{\label{fig:fscd-cartoon} A sketch of an antenna array large-volume CRES experiment. Electrons from $\beta$-decays are confined in a magnetic field using a set of trap coils. The cyclotron radiation produced by the motion of the trapped electrons can be detected by a surrounding antenna array to determine the electron energies. Measuring the energies of many electrons produces a $\beta$-decay spectrum.}
\end{figure}

The Project 8 collaboration is developing new methods for neutrino mass measurement based on Cyclotron Radiation Emission Spectroscopy (CRES) \cite{Monreal:2009za, Project8:2014ivu, Project8:2017nal, Project8:2022wqh}, with the goal of measuring the absolute scale of the neutrino mass with a 40 $\textrm{meV}/\textrm{c}^2$ sensitivity \cite{PhysRevC.103.065501, FORMAGGIO20211}. This sensitivity goal will require the development of two separate technical capabilities. First is the development of an atomic tritium source, which avoids significant spectral broadening due to molecular final states \cite{Bodine:2015sma}. Second is the technology for performing CRES in a multi-cubic-meter experimental volume with high combined detection and reconstruction efficiency, which is required in order to obtain sufficient event statistics near the tritium spectrum endpoint.

\begin{figure}[htbp]
\centering
\includegraphics[width=.4\textwidth]{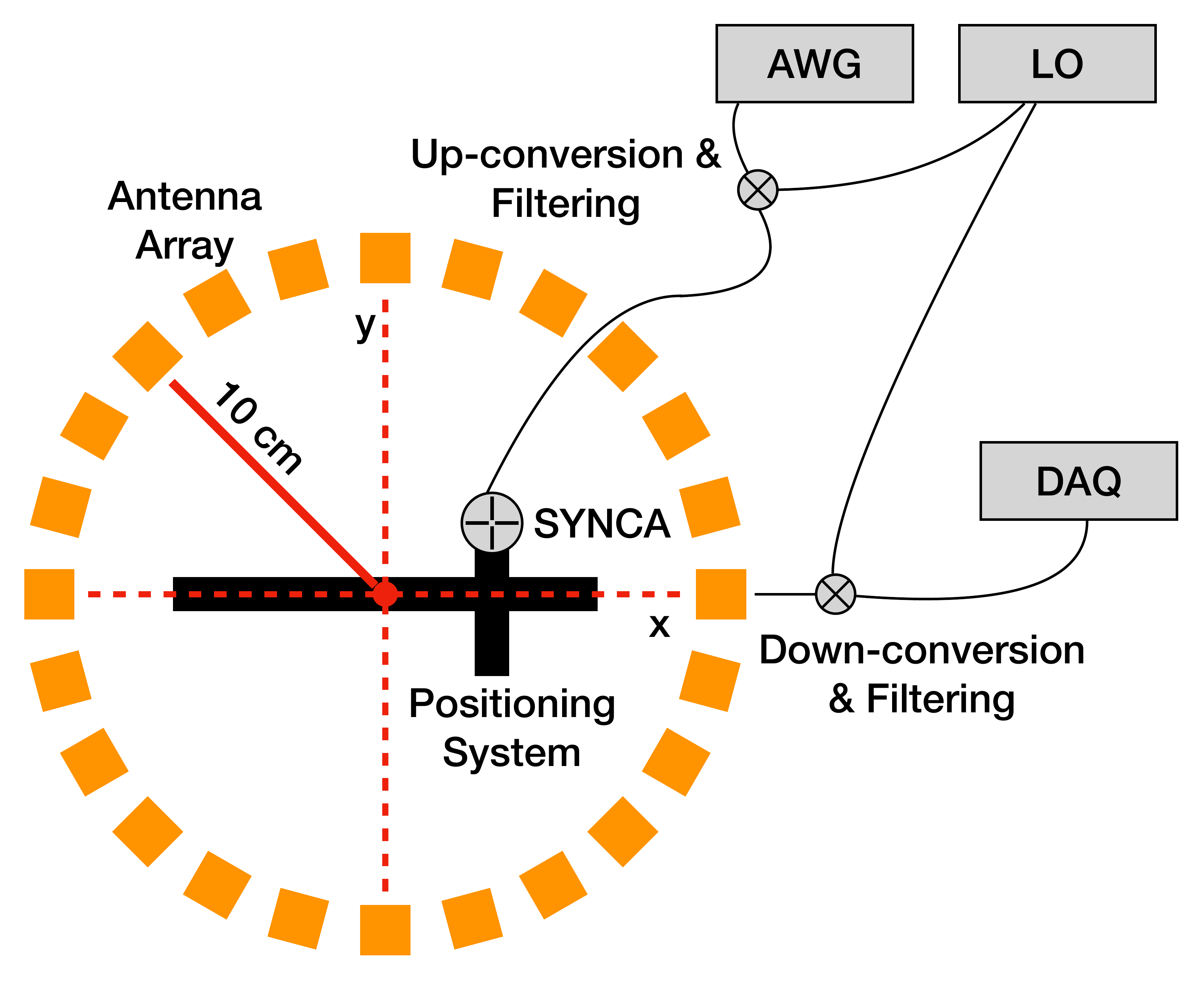}
\qquad
\caption{\label{fig:testbed-cartoon} A schematic of the antenna array test stand. The circular antenna array has a radius of 10~cm with 60 independent channels (limited number shown for clarity). The test stand includes an arbitrary waveform generator (AWG), local oscillator (LO), and data acquisition (DAQ) hardware. Finally, a specialized Synthetic Cyclotron Antenna (SYNCA) is used to inject signals to test the antenna array.}
\end{figure}

One approach for a large-volume CRES experiment is to use an array of antennas, which surrounds a volume of tritium gas, to detect the cyclotron radiation produced by the $\beta$-decay electrons when they are trapped in a background magnetic field using a set of magnetic trapping coils (see Figure \ref{fig:fscd-cartoon}). Project 8 has developed a conceptual experiment design to study the feasibility of this approach. The design consists of a single circular array of antennas with a radius of 10~cm and 60 independent channels positioned around the center of the magnetic trap. The motivation behind this antenna array design is to first develop an understanding of the antenna array approach to CRES with a small scale experiment before attempting to scale the technique to large volumes by using multiple antenna rings to construct the full cylindrical array. The development of the antenna array approach to CRES has largely proceeded through simulations using the Locust software package \cite{Ashtari_Esfahani_2019, nb_thesis}, which is used to model the fields emitted by CRES events and predict the signals received by the surrounding antenna array. To validate these simulations, a dedicated test stand is being constructed to perform characterization measurements of the prototype antenna array developed by Project 8 (see Figure \ref{fig:testbed-cartoon}) and benchmark signal reconstruction methods using a specially designed transmitting calibration probe antenna.

We call this probe antenna the Synthetic Cyclotron Antenna or SYNCA. The SYNCA is a novel antenna design that mimics the cyclotron radiation generated by individual charged particles trapped in a magnetic field, which will be used in the antenna test stand to perform characterization measurements, simulation validation, and reconstruction benchmarking. This paper provides an overview of the design, construction, and characterization measurements of the SYNCA performed in preparation for its usage as a transmitting calibration probe. 

In Section 2 we provide a description of the cyclotron radiation field characteristics that we recreate with the SYNCA. In Section 3 we give an overview of the simulations performed to develop an antenna design that mimics the characteristics of cyclotron radiation. In Section 4 we outline characterization measurements to validate that the fields generated by the SYNCA match simulation, and finally in Section 5 we demonstrate an application of the SYNCA to test phased array reconstruction techniques on the bench-top.

\section{Cyclotron Radiation Phenomenology}
\label{sec:pheno}

\begin{figure}[htbp]
\centering
\includegraphics[width=.5\textwidth]{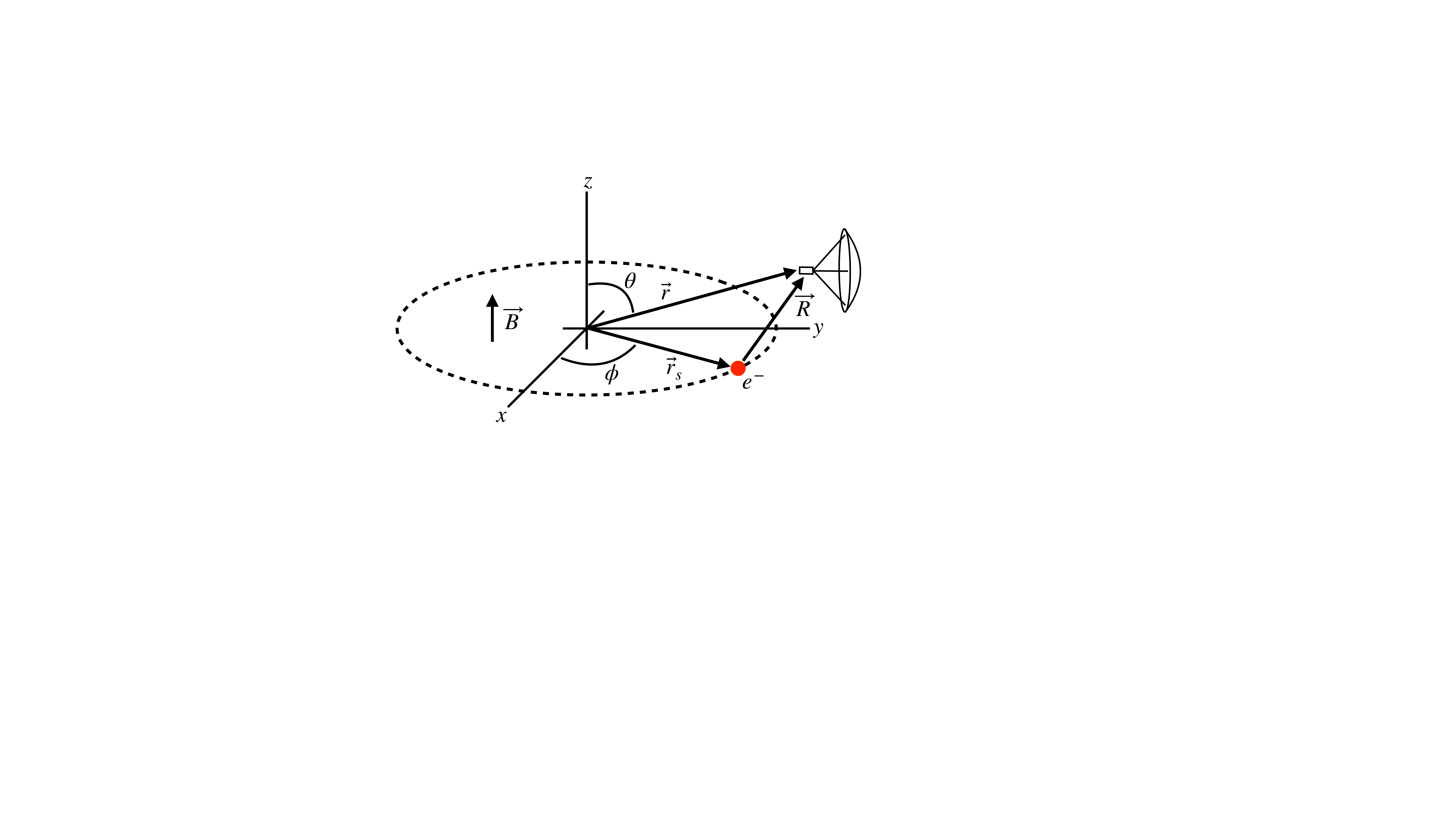}
\qquad
\caption{\label{fig:physics-situation} An electron (red dot) performing cyclotron motion in the x-y plane. The resulting cyclotron radiation is observed by an antenna located at the field point of interest. }
\end{figure}

To understand the cyclotron radiation phenomenology that the SYNCA should mimic, we consider a charged particle moving at relativistic speed in the presence of an external magnetic field (see Figure \ref{fig:physics-situation}). In the special case we shall examine, the entirety of the electron's momentum is directed perpendicular to the magnetic field; therefore, the trajectory of the electron is confined to the cyclotron orbit plane. Because the momentum vector is oriented perpendicular to the magnetic field, electrons with these trajectories are said to have pitch angles of $90^\circ$. 

The cyclotron radiation fields generated by this circular trajectory are those which we aim to reproduce with the SYNCA. We can describe the electromagnetic (EM) fields using the Li\'{e}nard-Wiechert equations \cite{jackson_classical_1999, nb_thesis}, which in non-covariant form express the electric field as
\begin{equation}
    \label{eq:lw-e}
      \vec{E}  =e\left[\frac{\hat{n}-\vec{\beta}}{\gamma^2(1-\vec{\beta}\cdot\hat{n})^3|\vec{R}|^2}\right]_{t_\textrm{r}}
      +\frac{e}{c}\left[\frac{\hat{n}\times[(\hat{n}-\vec{\beta})\times\dot{\vec{\beta}}]}{(1-\vec{\beta}\cdot\hat{n})^3|\vec{R}|}\right]_{t_\textrm{r}},
\end{equation}
where $e$ is the particle's charge, $\hat n = (\vec{r}-\vec{r_s})/|\vec{r}-\vec{r_s}|$ is the unit vector pointing from the electron to the field measurement point, $\vec\beta=\dot{\vec{r_s}}/c$ is the velocity of the particle divided by the speed of light, and $\gamma$ is the relativistic Lorentz factor. The equation is meant to be evaluated at the retarded time as indicated by $t_\textrm{r}=t-|\vec{R}|/c$, which accounts for the time delay due to the finite speed of light between the point where the field was emitted and the point where the field is detected.

In Appendix \ref{sec:app-1} we show that the electric fields received by an antenna positioned in the plane of the cyclotron orbit and located at a distance that is much greater than the radius of the cyclotron orbit is a sum of a radially and azimuthally polarized fields. For an electron with a kinetic energy near the 18.6~keV Q-value of the tritium $\beta$-decay \cite{Bodine:2015sma} moving in a 1~T magnetic field, we show that the time-averaged amplitude of the radially polarized electric field is a factor of 8 smaller than the azimuthally polarized component. Additionally, we make the assumption that the receiving antenna is designed to have a gain that is significantly larger for the azimuthal electric fields, which is true for the antenna array in the test stand mentioned in Section \ref{sec:intro}. Therefore, we can approximate the electric fields from Equation \ref{eq:lw-e} as purely azimuthally or $\phi$-polarized. The simplified expression for the electric field received by an antenna becomes
\begin{equation}
    \label{eq:lw-e-phi-simple}
    \vec{E}=E_\phi\hat{\phi} = \frac{e\frac{r_c\omega_c}{c}}{4r_{a}r_c}\left[\frac{\frac{r_c\omega_c}{c}-\cos{\omega_ct}-\frac{4r_c}{ r_{a}}\sin{\omega_ct}}{(1-\frac{r_c\omega_c}{c}\cos{\omega_ct})^3}\right]_{t_r}\hat{\phi},
\end{equation}
where the radius of the cyclotron orbit is called $r_c$, the cyclotron frequency is called $\omega_c$, and the radial position of the receiving antenna is called $r_a$. Equation \ref{eq:lw-e-phi-simple} has been evaluated in the non-relativistic limit where $\gamma\simeq1$, which is justified by the fact that $|\vec{\beta}|\simeq\frac{c}{4}$ for an electron with an 18.6~keV kinetic energy in a 1~T magnetic field.

The rather complicated expression for the electric field can be simplified using Fourier analysis. In Appendix \ref{sec:app-2} we show that Equation \ref{eq:lw-e-phi-simple} can be represented by a sum of pure sinusoids with frequencies that are harmonics of the main cyclotron carrier frequency,
\begin{equation}
    \label{eq:lw-e-phi-final}
    \vec{E} = \frac{e\frac{r_c\omega_c}{c}}{4r_{a}r_c}\left[\sum_{k=0}^7{A_k e^{i\omega_k t}}\right]_{t_r}\hat{\phi}.
\end{equation}
The frequencies ($\omega_k=k\omega_c$) are uniformly spaced by integer values of the electron's cyclotron frequency of $25.898$~GHz and include a constant term from the electron's coulomb field at zero frequency. The relative amplitudes ($A_k$) are dimensionless complex numbers which encode the relative amplitudes of the harmonics and the overall starting phase of the cyclotron motion. Because magnitude of the relative amplitudes exponentially decreases for higher harmonics, it is usually sufficient to consider only the terms up to $k=4$ where the relative amplitude of the hamonics has decresed from the main carrier by a factor of approximately 100. However, for completeness we include harmonics up to 7th order in Equation \ref{eq:lw-e-phi-final}. The range of frequencies to which the receiving antenna array in the antenna test stand is sensitive is defined by the antenna's transfer function. The receptive bandwidth for the antennas used in the test stand is a range of frequencies with a bandwidth on the order of a few GHz centered around the main cyclotron carrier frequency of 25.898~GHz. Therefore, the higher order harmonics as well as the zero frequency term can be ignored when considering only the signals that will be received by the antenna array.

\begin{figure}[h]
    \centering
    \begin{subfigure}[b]{0.48\textwidth}
        \centering
        \includegraphics[width=.8\textwidth]{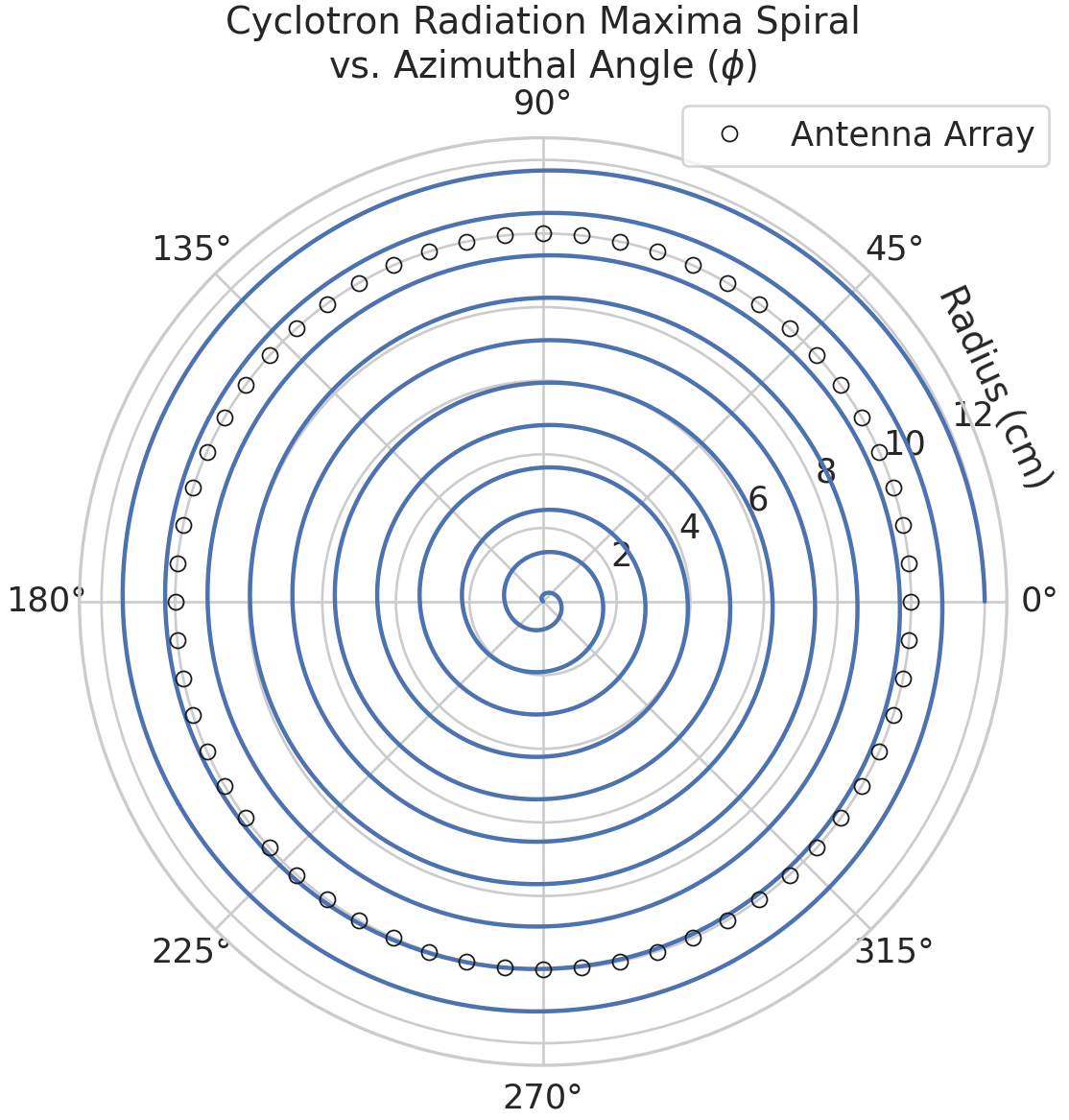}
        \caption{\label{fig:arch-spiral}}
    \end{subfigure}
    \hfill
    \begin{subfigure}[b]{0.48\textwidth}
        \centering
        \includegraphics[width=.8\textwidth]{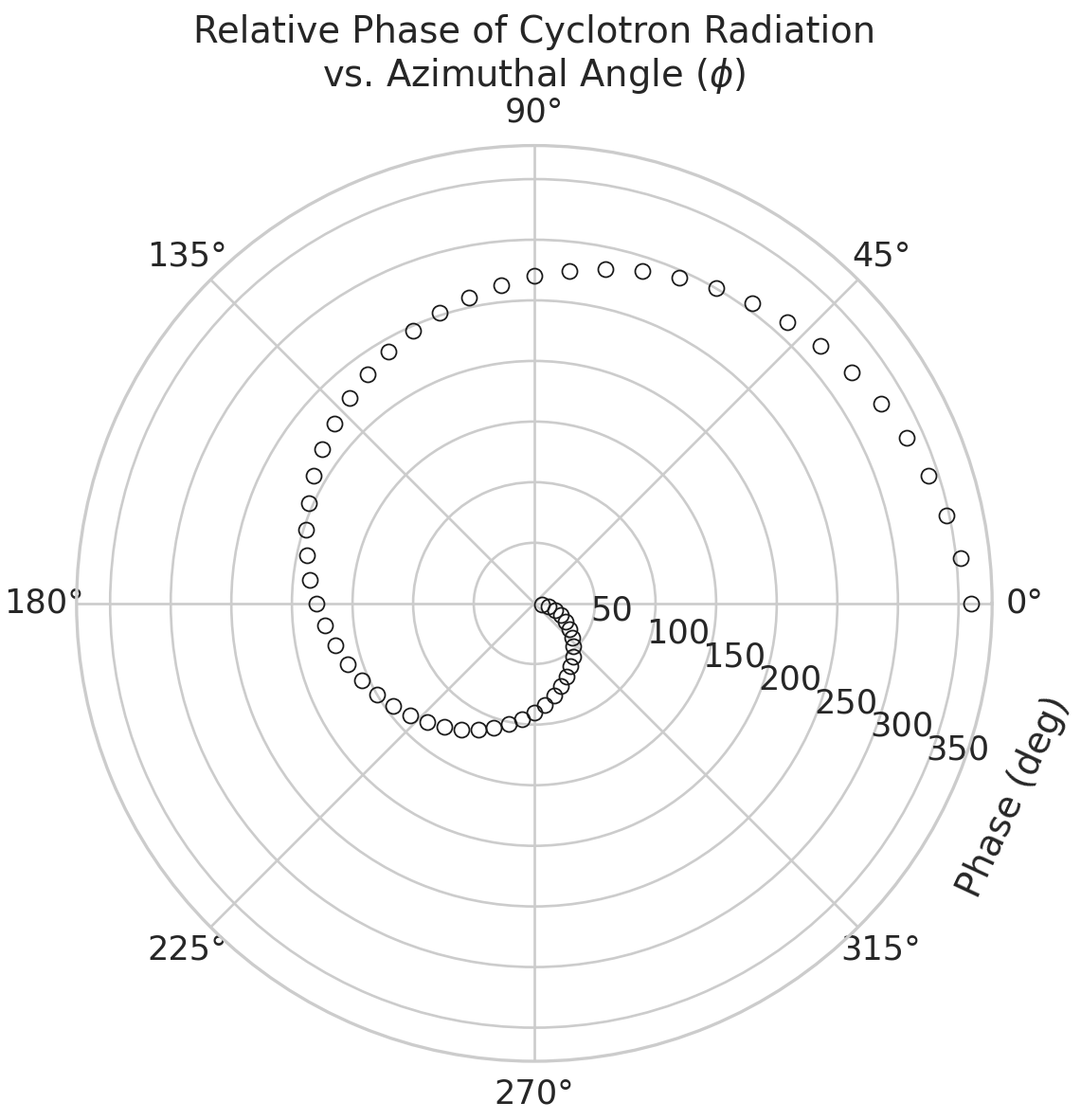}
        \caption{\label{fig:array-sprial}}
    \end{subfigure}
    \hfill
    \caption{ The amplitude maxima of the cyclotron radiation form an Archimedean spiral as the radiation propagates outward from the cyclotron orbit center (a). A circular antenna array located at a fixed radius from the orbit center will receive electric fields with equal magnitude in each of its channels, but the phase of the electric field incident on each array channel will be linearly out of phase from its neighbor antennas by an amount equal to the angular separation of the two channels (b).}
    \qquad
\end{figure}

Considering only the 1st order harmonic term from Equation \ref{eq:lw-e-phi-final}, which represents the portion of the electric field that will be detected by the array, and evaluating this at the retarded time we obtain the following for the $\phi$-polarized electric fields
\begin{equation}
    \label{eq:lw-e-phi}
    E_\phi \propto \cos{\left(\omega_c\left(t - |\vec{R}|/c\right)-\Delta\right)},
\end{equation}
where the arbitrary phase $\Delta$ is defined by $A_k=|A_k|e^{i\Delta}$. We are interested in the characteristics of the amplitude of the electric field as a function of the radial distance component ($|\vec{R}|$) of the retarded time. In particular, the maximum of $E_\phi$ occurs when the argument of the cosine function is equal $n\pi$ where $n\in\{0,\pm2,\pm4,...\}$; however, the solutions where $n$ is negative can be discarded since they represent unphysical negative overall phases. Applying this condition to Equation \ref{eq:lw-e-phi} gives a condition on the radial position of the maximum of $E_\phi$
\begin{subequations}
\label{eq:ephi_max}
  \begin{align}
      \omega_c (t-|\vec{R}|/c)-\Delta & = n\pi,\\
      |\vec{R}| & = \frac{c}{\omega_c}\left((\omega_c t-\Delta)-n\pi\right),
  \end{align}
\end{subequations}
which is a function of time in the frame of the moving electron ($t$). Equation \ref{eq:ephi_max} can be further simplified by noticing that the azimuthal position of the electron ($\phi_e(t)$) as a function of time is defined by $\phi_e(t)=\omega_c t - \Delta$ which reduces Equation \ref{eq:ephi_max} to
\begin{equation}
\label{eq:spiral}
    |\vec R| = \frac{c}{\omega_c}(\phi_e(t)-n\pi).
\end{equation}
Equation \ref{eq:spiral} represents an archimedian spiral which is formed when plotting the amplitude of $E_\phi$ in the x-y plane. The solution where $n=0$ represents the leading edge of the radiation spiral which propagates outward from the electron at the speed of light. The additional solutions for $n>0$ represent the persistent spiral at radii inside the leading edge of the radiated fields that have not yet been detected by the receiver at the current time. In Figure \ref{fig:arch-spiral} we show the expected spiral pattern for the maxima of the cyclotron radiation. 

In particular, we note that for the circular array geometry of the test stand, depicted as the series of circles in Figure \ref{fig:arch-spiral}, each antenna receives a linearly polarized wave with a phase offset that corresponds to the azimuthal angle for that antenna element. Therefore, as we show in Figure \ref{fig:array-sprial}, when the relative phase of the received signal is plotted as a function of the recieving antenna's azimuthal position the result is also an Archimedean spiral. 

Based on these analytical calculations we can characterize the magnitude, polarization, and phase of the signals received by the antenna array using three criteria. These criteria are the basis of comparison for the radiation produced by the SYNCA and cyclotron radiation emitted by electrons and will be used to evaluate the performance of antenna designs. The criteria are:
\begin{enumerate}
    \item Electric fields that are $\phi$-polarized near $\theta=90^\circ$
    \item Uniform time-averaged electric field magnitudes around the circumference of a circle centered on the antenna
    \item Electric fields whose phase is equal to the azimuthal angle at the point of measurement plus a constant
\end{enumerate}

The Locust simulation package \cite{Ashtari_Esfahani_2019} can be used to directly simulate the EM fields generated by electrons performing cyclotron motion to validate the analytical calculations. Locust simulates the EM fields by first calculating the trajectory of the electrons in the magnetic trap using the Kassiopeia software package \cite{Furse_2017}. The trajectory can then be used to solve for the EM fields using the Li\'{e}nard-Wiechert equations directly with no approximations. The resulting electric field solutions drive a receiving antenna by convolving the time-domain fields with the finite-impulse response filter of the antenna or they can be examined directly to study the field characteristics that the SYNCA must reproduce. In the next section we compare the radiation field patterns for electrons simulated with Locust to patterns from a SYNCA antenna design.

\section{Design of the SYNCA}
\label{sec:synca_design}
\begin{figure}[h]
    \centering
    \includegraphics[width=0.7\textwidth]{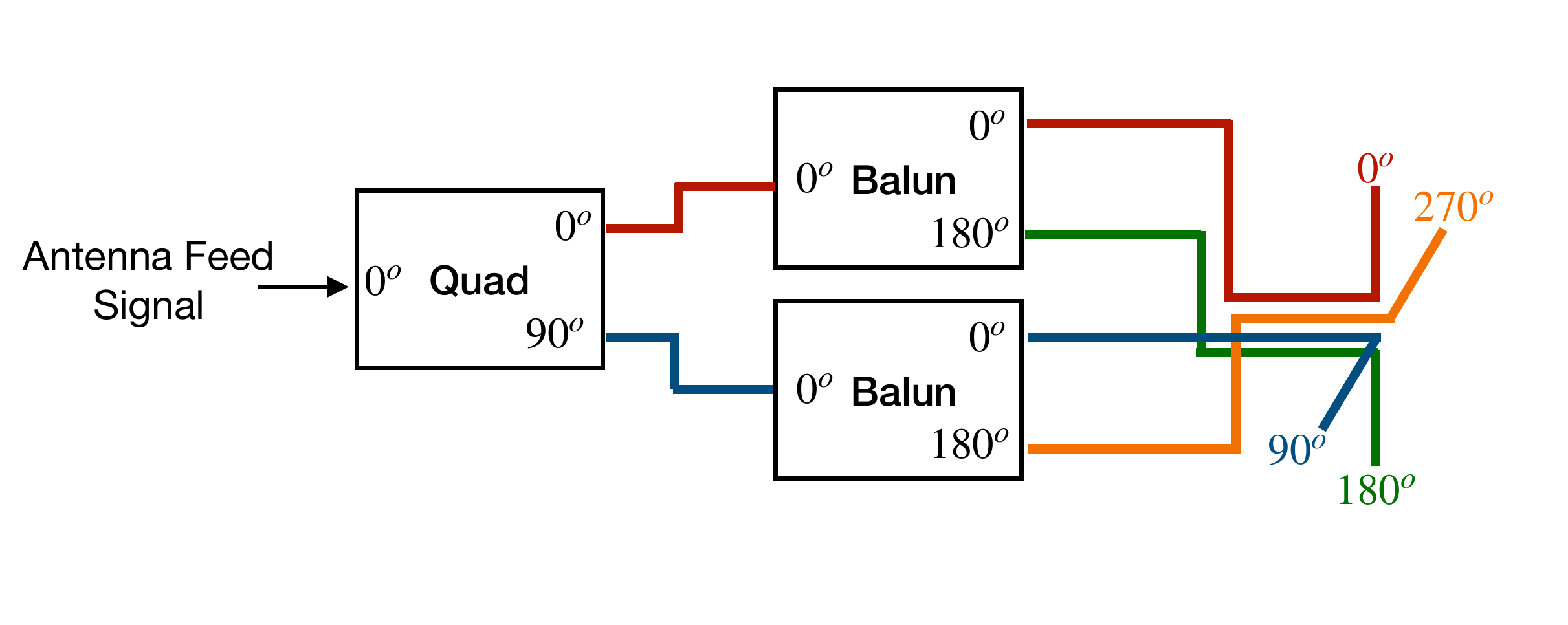}
    \caption{\label{fig:cross-dipole} An idealized crossed-dipole antenna consists of two electric dipole antennas oriented perpendicular to each other and is fed with four signals with a quadrature phase relationship. An example antenna feed circuit is shown which is composed of a chained combination of a quadrature hybrid-coupler (Quad) and two baluns. }
\end{figure}

One potential SYNCA design is the crossed-dipole antenna \cite{balanis2011modern}. A crossed-dipole antenna consists of two dipole antennas, one of which is rotated $90^\circ$ with respect to the other, which are fed with signals that are out of phase from the opposite dipole by $90^\circ$ (see Figure \ref{fig:cross-dipole}). 
This arrangement causes the signals fed to each arm of the dipole to be out of phase from each of the neighboring arms by $90^\circ$, which mirrors the spatial phase relationship of cyclotron radiation fields.

A potential drawback of this design is that standard crossed-dipole antennas do not radiate uniform electric fields near the $\theta=\pi/2$ plane. Typical crossed-dipole antennas use dipole arm lengths equal to $\lambda/4$ or larger \cite{balanis2011modern}, where $\lambda$ is the wavelength at the desired operating frequency. Such large arm lengths cause the electric field magnitude to vary significantly around the circumference of the antenna. However, making the antenna electrically small by shrinking the arm length can improve the antenna pattern uniformity.

In general, the criterion for an electrically small antenna is that the largest dimension of the antenna ($D$) obey $D\lesssim\lambda/10$ \cite{balanis2015antenna}. In our application, we are attempting to mimic the cyclotron radiation emitted by electrons produced from tritium $\beta$-decay with energies near the spectrum endpoint. For a background magnetic field of 1~T, the corresponding cyclotron frequency of tritium endpoint electrons is approximately 26~GHz. Therefore, the electrically small condition would require that the largest dimension of the crossed-dipole antenna be smaller than 1.2~mm.

\begin{figure}[h]
\centering
\includegraphics[width=.5\textwidth]{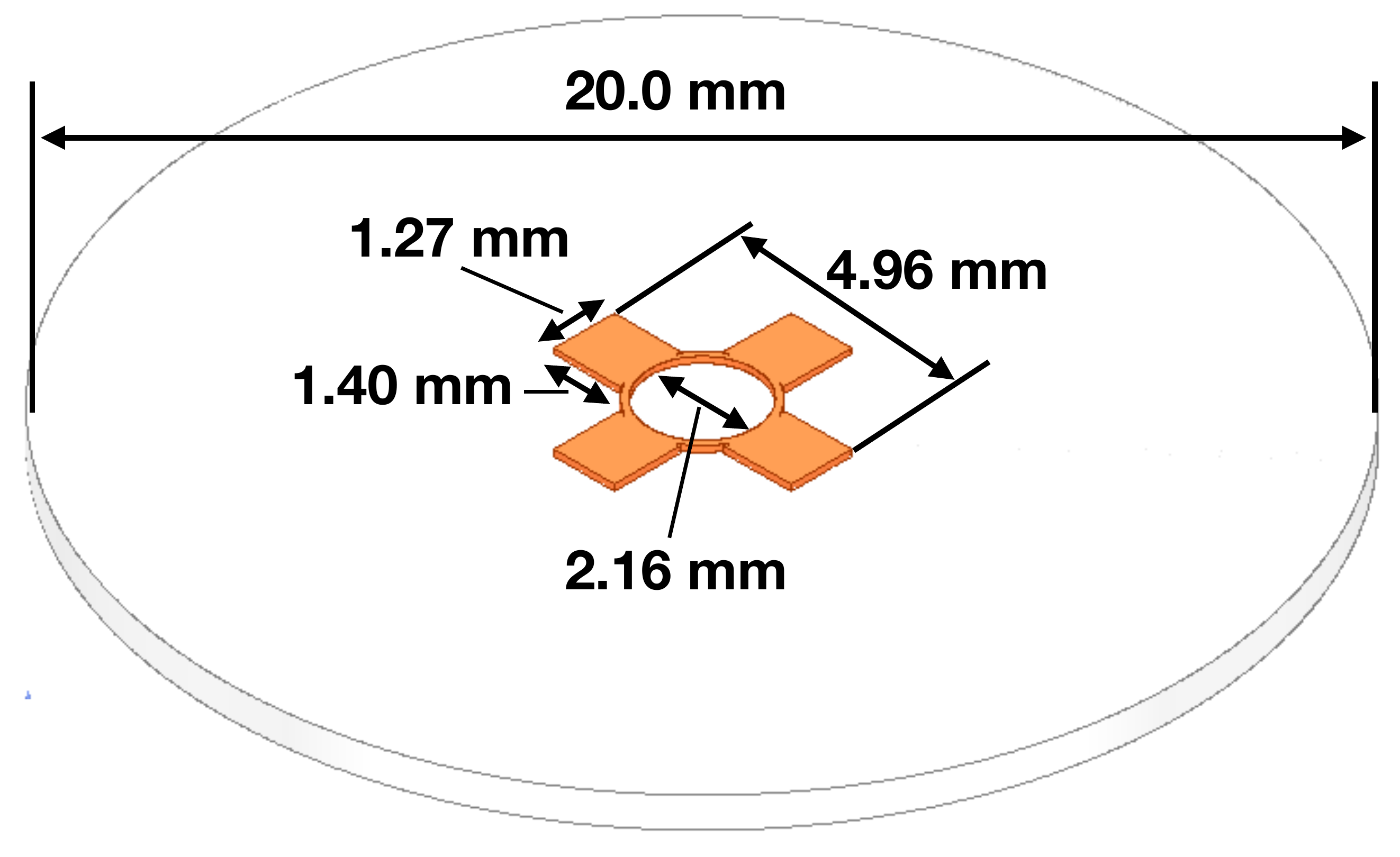}
\qquad
\caption{\label{fig:cross-dipole-pcb-model} A model of the PCB crossed-dipole antenna with dimensions. The design has an inside diameter of 2.16~mm for the central circular trace, which is 0.13~mm wide. The dipole arms each have a width of 1.27~mm and protrude beyond the circular trace by 1.40~mm, which gives an overall width of 4.96~mm for the length of the antenna PCB trace from end-to-end. The overall size of the antenna is 20.0~mm the majority of which is the PCB dielectric material. This design was observed in simulation to maintain the field characteristics of the idealized crossed-dipole while being simpler to fabricate due to the increased size of the antenna.}
\end{figure}

A crossed-dipole antenna with an overall size of 1.2~mm is challenging to fabricate due to the small dimensions of the dipole arms that, in practice, are fragile and unsuitable for use as a calibration probe. To mitigate some of the challenges with the fabrication of such a small antenna, a variant crossed-dipole antenna design using printed circuit board (PCB) technology (see Figure \ref{fig:cross-dipole-pcb-model}) was developed in partnership with an antenna prototyping company, Field Theory Consulting \footnote{https://fieldtheoryinc.com/}. 

The PCB crossed-dipole design uses four rectangular pads to represent the dipole arms, which are connected by a thin circular trace. The circular trace both adds mechanical stability to the antenna and improves the azimuthal uniformity of the electric fields compared to a more standard crossed-dipole geometry. Futhermore, the circular trace allows for a greater separation between dipole arms than standard crossed-dipoles, which is required to accommodate the coaxial connections to each pad. The pads each contain a through-hole solder joint to connect coaxial transmission lines using hand soldering. The antenna PCB has no ground plane on the bottom layer as this was observed in simulation to significantly distort the radiation pattern in the plane of the PCB. The only ground planes present in the model are the outer conductors of the four coaxial transmission lines which feed the antenna. These are left unterminated on the bottom of the PCB dielectric material.

\begin{figure}[h]
    \centering
    \includegraphics[width=1.\textwidth]{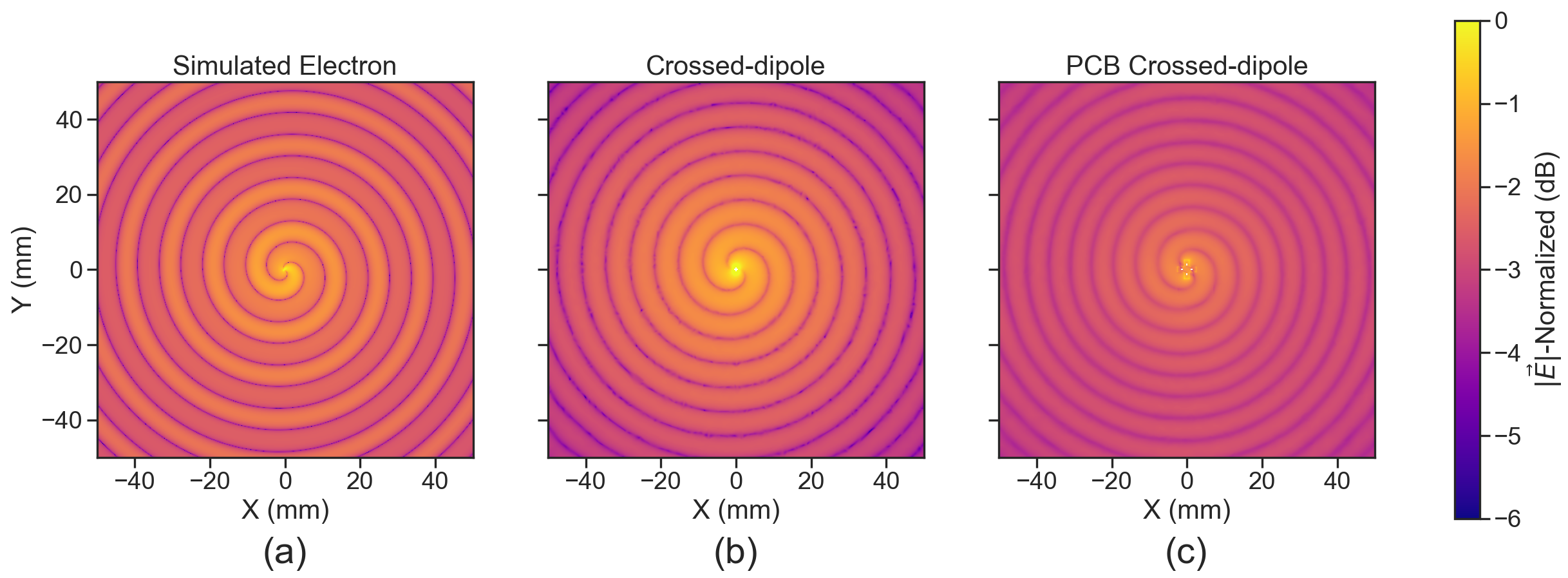}
    \caption{A comparison of the electric field magnitudes, normalized by the maximum value of the electric field in each simulation, plotted on a 10~cm square to visualize the Archimedean spirals formed by the electron (a), the crossed-dipole antenna (b), and a PCB crossed-dipole antenna (c). The matching patterns indicate that the electric fields have similar phase characteristics. These images were generated using Locust simulations for the electron and ANSYS HFSS for both antennas.}
    \label{fig:spiral-comparison}
\end{figure}

The antenna design development utilized a combination of Locust electron simulations and antenna simulations using ANSYS HFSS \cite{hfss}, a commercial finite-element electromagnetic simulation software. Two antenna designs were simulated: an idealized electrically small crossed-dipole antenna with an arm length of 0.40~mm and an arm separation of 0.05~mm, as well as a PCB crossed-dipole antenna with the dimensions shown in Figure \ref{fig:cross-dipole-pcb-model}. Plotting the magnitude of the electric fields generated by the antennas across a 10~cm square located in the same plane as the respective antennas reveals the expected cyclotron spiral pattern (see Figure \ref{fig:spiral-comparison}) which closely matches the prediction for simulated electrons. The spiral pattern demonstrates that the electric fields have the appropriate phases to mimic cyclotron radiation, which fulfills SYNCA criterion 3 identified in Section \ref{sec:pheno}.

\begin{figure}[h]
    \centering
    \includegraphics[width=1.\textwidth]{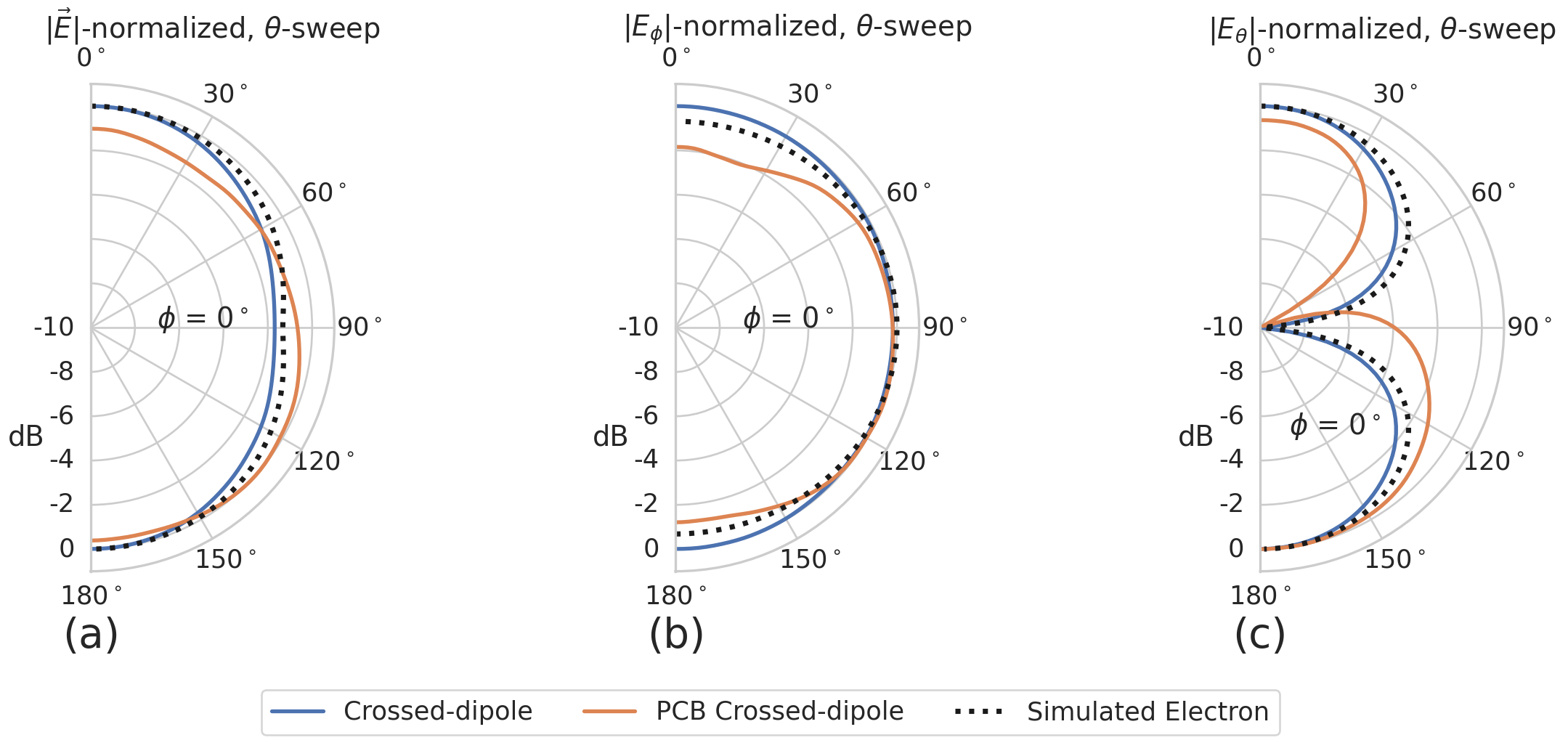}
    \caption{A comparison of the normalized electric field magnitudes for the ideal crossed-dipole, PCB crossed-dipole, and a simulated electron as a function of the polar angle ($\theta$). (a) Shows the total electric field, (b) shows the $\phi$-polarized electric field component, and (c) shows the $\theta$-polarized electric field component. These images were generated using Locust simulations for the electron and ANSYS HFSS for both antennas.} 
    \label{fig:field-comparison-theta}
\end{figure}

As we can see from Figure \ref{fig:field-comparison-theta}, the crossed-dipole antenna, which uses an idealized geometry, exhibits good agreement with simulation. The antenna has a maximum deviation from a simulated electron of approximately 0.5 dB in the total electric field, 1 dB for the $\phi$-polarized electric field and 1 dB for the $\theta$-polarized electric field.

In comparison, the pattern of the PCB crossed-dipole antenna, because the simulation incorporates the geometry of the coax transmission lines, exhibits some distortion from the idealized cross-dipole simulations. The vertically oriented ground planes of the coax lines introduce more $\theta$-polarized electric fields than are observed for simulated electrons near $\theta=90^\circ$. The significant $\theta$-polarized field minimum is still present but shifted to approximately $\theta=65^\circ$.
The $\theta$-polarized field deviations of the PCB crossed-dipole antenna should not greatly impact the performance of the antenna because the receiving antenna array is primarily $\phi$-polarized. Therefore deviations in the $\theta$-polarized fields will be suppressed due to the polarization mismatch. More importantly, the $\phi$-polarized electric field pattern generated by the PCB crossed-dipole closely matches simulated electrons across the polar angle range of $50^\circ<\theta<150^\circ$. In this region the PCB crossed-dipole differs by less than $0.5$~dB from simulated electrons. This range greatly exceeds the beamwidth of the receiving antenna array which is designed to be most sensitive to fields produced near $\theta=90^\circ$. Therefore, we conclude that the PCB crossed-dipole antenna generates a $\phi$-polarized radiation pattern that fulfills SYNCA criterion 1 from Section \ref{sec:pheno}.

\begin{figure}[h]
    \centering
    \includegraphics[width=0.45\textwidth]{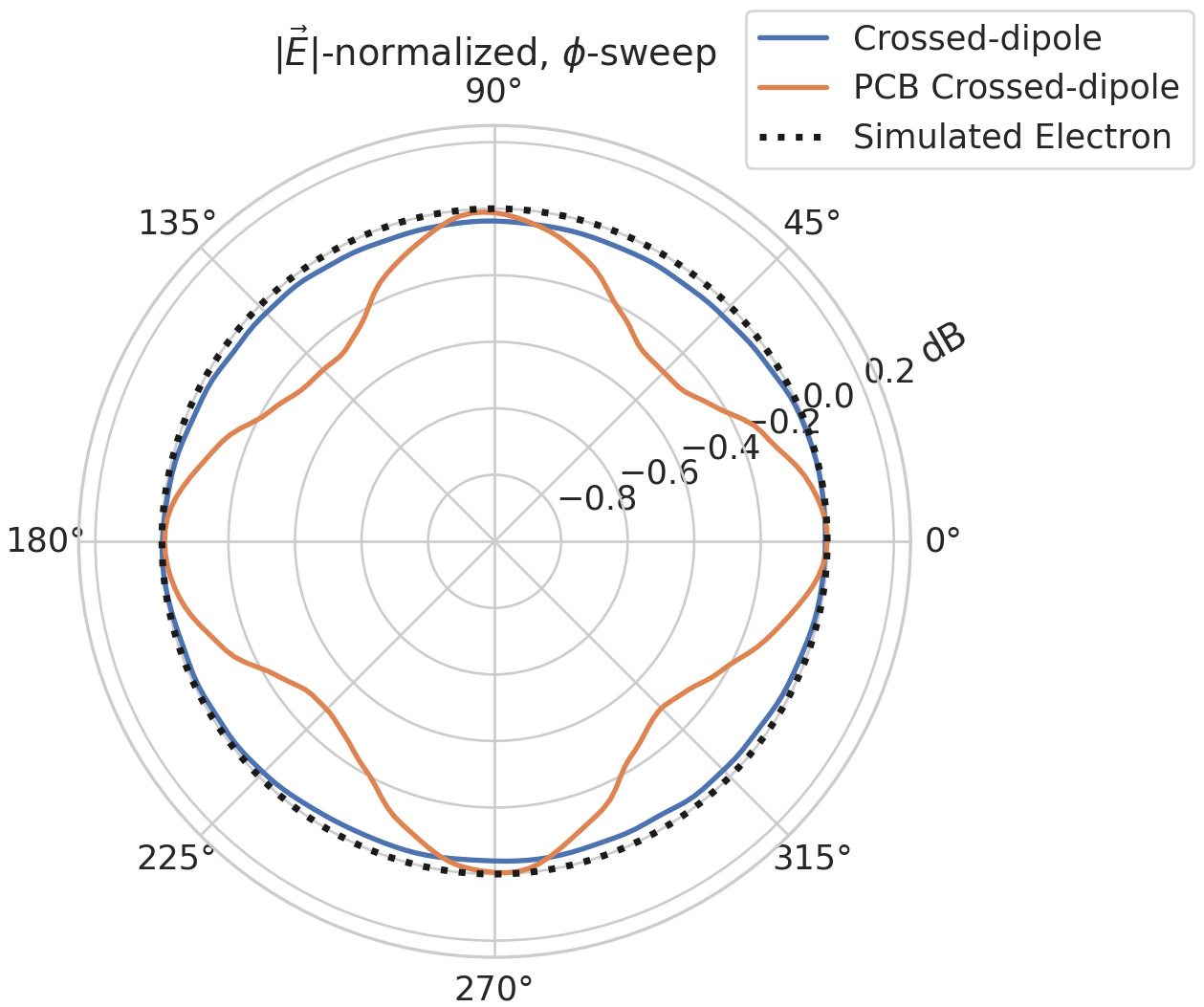}
    \caption{A comparison of the normalized electric field magnitudes for the crossed-dipole, PCB crossed-dipole, and a simulated electron as a function of the azimuthal angle ($\phi$) evaluated at $\theta=90^\circ$. This image was generated using Locust simulations for the electron and ANSYS HFSS for both antennas.}
    \label{fig:field-comparison-phi}
\end{figure}

The final SYNCA criterion is related to the uniformity of the electric fields when measured azimuthally around the antenna. As we saw for real electrons in Section \ref{sec:pheno} it is expected that the magnitude of the electric field be completely uniform as a function of the azimuthal angle due to the symmetry of the cyclotron orbit. In Figure \ref{fig:field-comparison-phi} we plot the total electric field as a function of azimuthal angle for an electron, the crossed-dipole antenna, and the PCB crossed-dipole antenna. 
The crossed-dipole antenna exhibits perfect uniformity around the azimuthal angle, whereas the PCB crossed-dipole has a small periodic deviation with a maximum difference of 0.3~dB caused by the coaxial transmission lines below the PCB. Such a small deviation from uniformity is acceptable since it is smaller than the expected variation in uniformity caused by imperfections in the antenna fabrication process, which modifies the antenna shape in an uncontrolled manner by introducing solder blobs with a typical size of a few tenths of a millimeter on the dipole arms (see Figure \ref{fig:prototype-antenna}). Additionally, the SYNCA will be separately calibrated to account for azimuthal differences in the electric field magnitude. Therefore we see from the simulated performance of the PCB crossed-dipole antenna that this antenna design meets all three of the SYNCA criteria.

\section{Characterization of the SYNCA}

\begin{figure}[h]
    \centering
    \begin{subfigure}[b]{0.48\textwidth}
        \centering
        \includegraphics[width=1\textwidth]{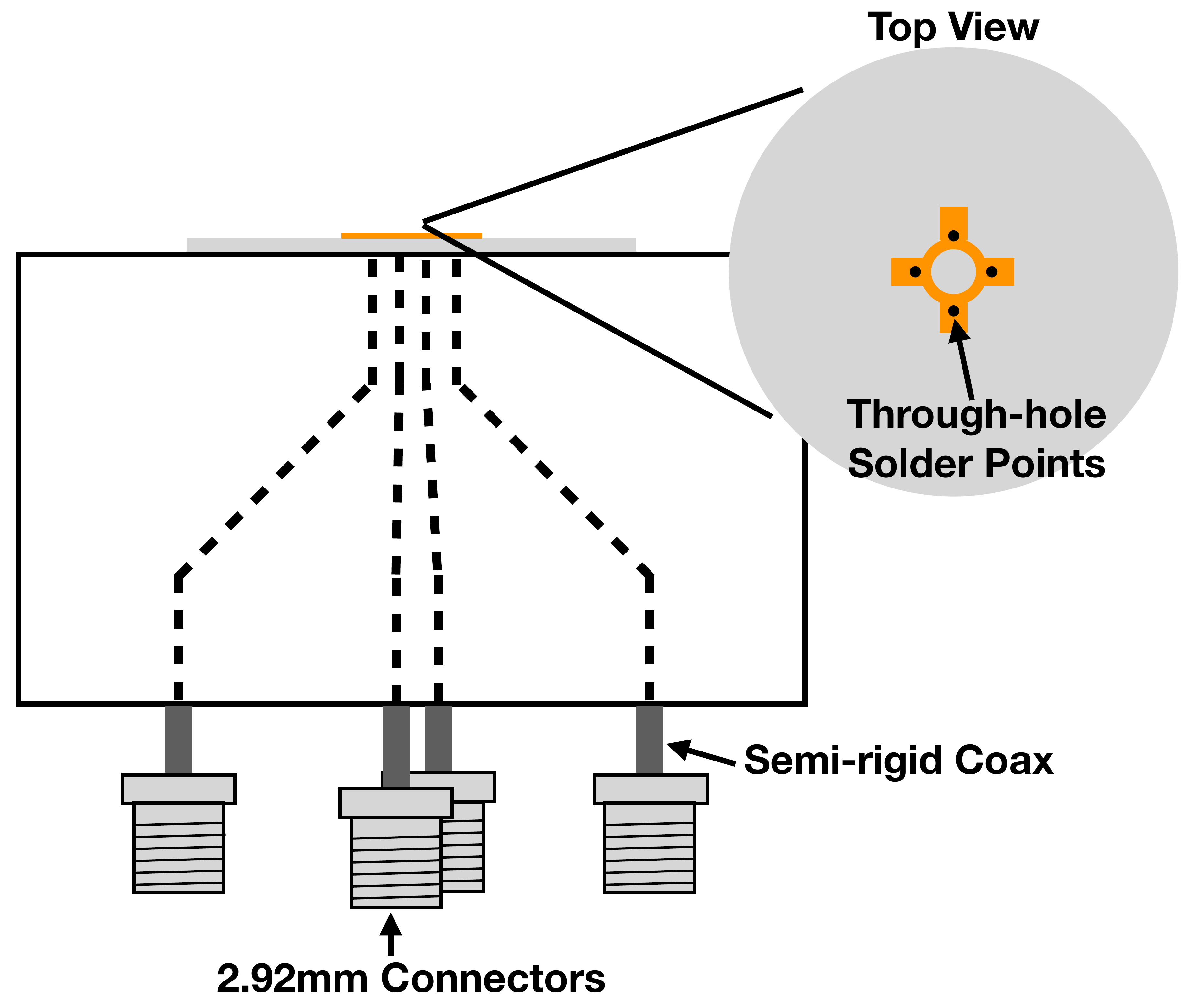}
        \caption{\label{fig:prototype-antenna-photo}}
    \end{subfigure}
    \hfill
    \begin{subfigure}[b]{0.48\textwidth}
        \centering
        \includegraphics[width=0.75\textwidth]{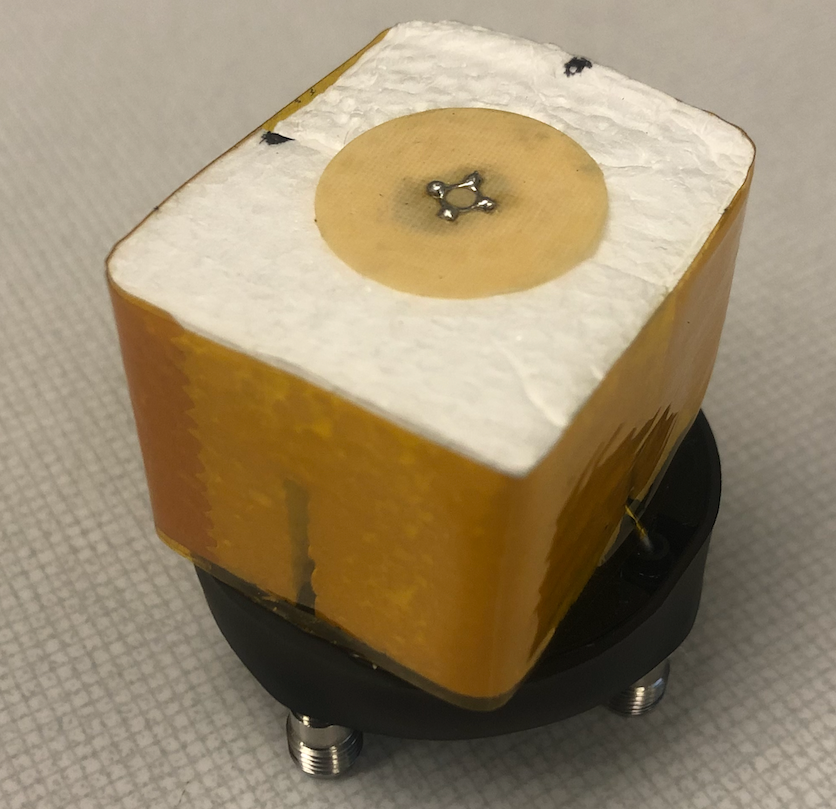}
        \caption{\label{fig:prototype-antenna-cartoon}}
    \end{subfigure}
    \hfill
    \caption{\label{fig:prototype-antenna} (a) A cartoon schematic which highlights the routing of the semi-rigid coax transmission lines. (b) A photograph of a SYNCA constructed using the modified crossed-dipole PCB antenna design. Visible in the photograph of the SYNCA are four blobs of solder which are an artifact of the SYNCA's hand-soldered construction. These solder blobs are the most significant deviation from the SYNCA design shown in Figure \ref{fig:cross-dipole-pcb-model} and are responsible for a significant fraction of the irregularities seen in the antenna pattern.}
    \qquad
\end{figure}

Two SYNCAs were manufactured using the PCB crossed-dipole design (see Figure \ref{fig:prototype-antenna}). The antenna PCB (Matrix Circuit Board Materials, MEGTRON 6) is connected to four 2.92~mm coaxial connectors (Fairview Microwave, SC5843) using semi-rigid coax (Fairview Microwave, FMBC002), which also physically support the antenna PCB. The antenna PCB consists only of two layers which correspond to the copper antenna trace and the PCB dielectric. Each coax line is connected to the associated dipole arm using through-hole soldering and phase matched to ensure that the electrical length of each of the transmission lines is identical at the operating frequency. The antenna PCB is further reinforced using custom cut polystyrene foam blocks, which have an electrical permittivity nearly identical to air. A custom 3D printed mount is included at the base of the antenna to support the coax connectors and to provide a sturdy mounting base.

\begin{figure}[t]
    \centering 
    \begin{subfigure}[b]{0.48\textwidth}
        \centering
        \includegraphics[width=\textwidth]{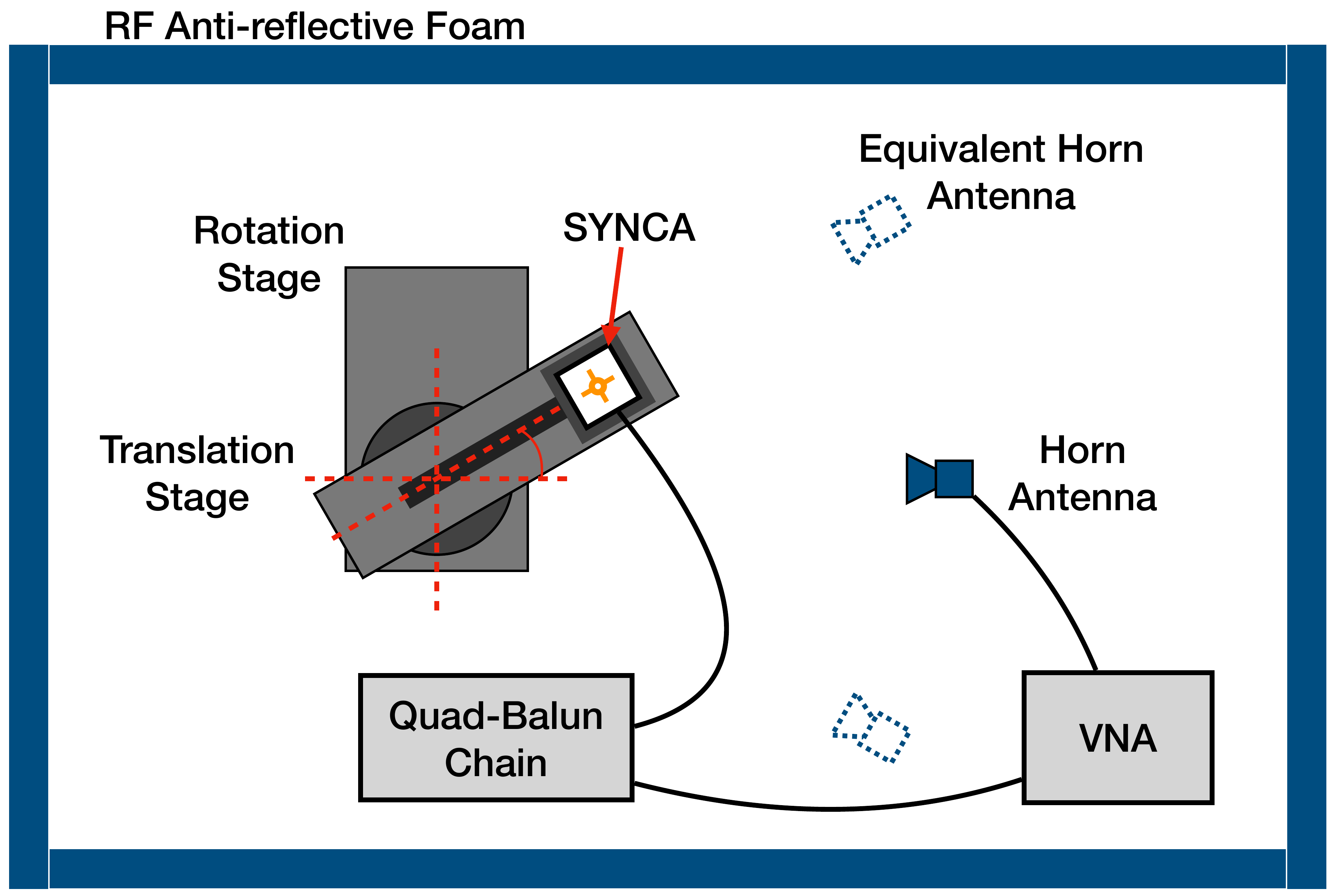}
        \caption{\label{fig:vna-schematic}}
    \end{subfigure}
    \hfill
    \begin{subfigure}[b]{0.45\textwidth}
        \centering
        \includegraphics[width=\textwidth]{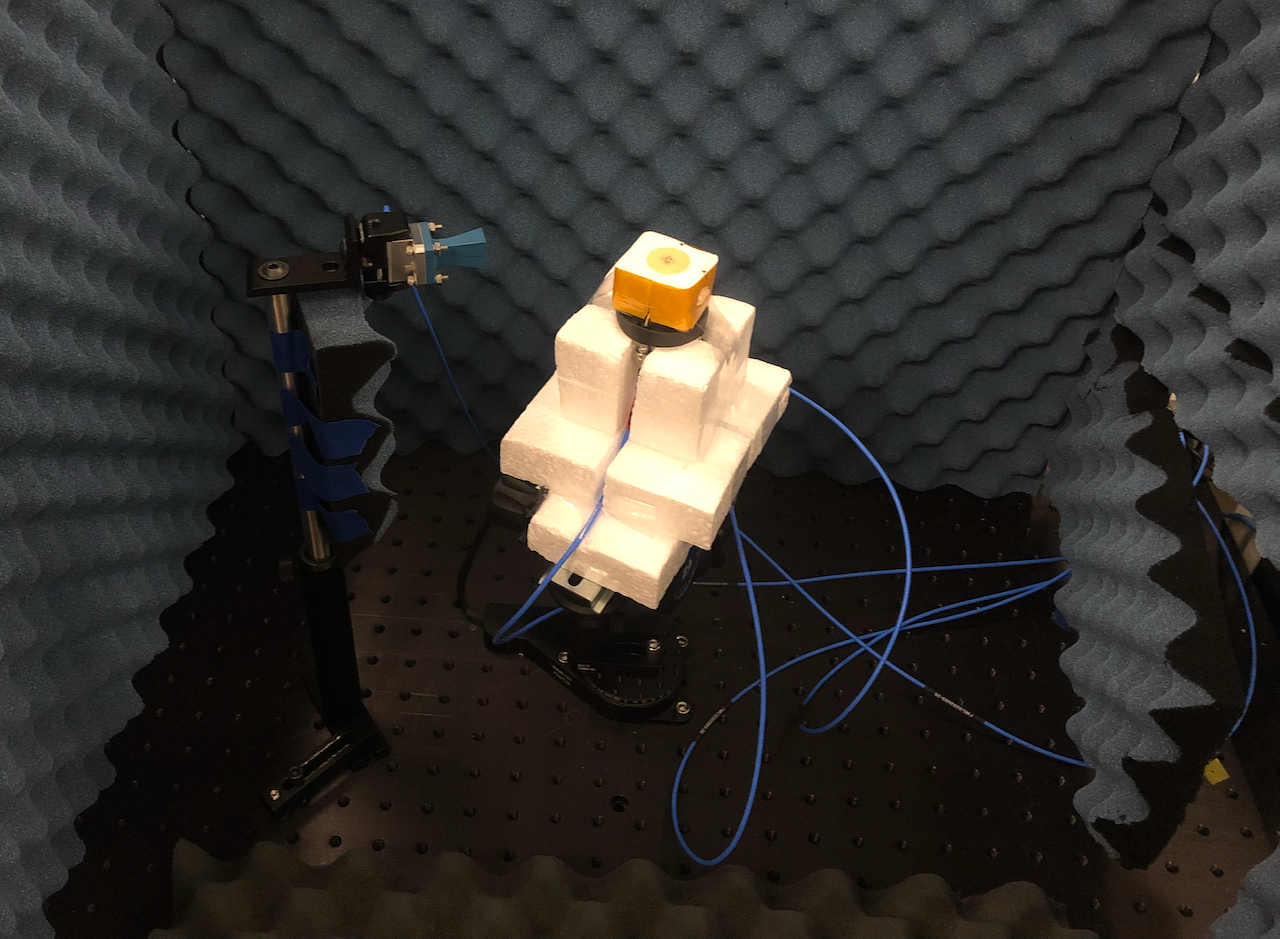}
        \caption{\label{fig:lab-meas-photo}}
    \end{subfigure}
    \hfill
    \caption{\label{fig:vna-meas-schematic} A schematic of the VNA characterization measurements (a). This setup allows for antenna gain and phase measurements across a full $360^\circ$ of azimuthal angles using a motorized rotation stage and control of the radial position of the SYNCA using a translation stage. A photo of the setup in the lab is shown in (b).}
    \qquad
\end{figure}

Characterization measurements were performed using a Vector Network Analyzer (VNA) to measure the electric field magnitude and phase radiated by the SYNCA to verify the radiation pattern (see Figure \ref{fig:vna-meas-schematic}). The VNA is connected to the SYNCA at one port through a hybrid-coupler whose outputs are connected to two baluns to generate the signals with the appropriate phases to feed the SYNCA (see Figure \ref{fig:cross-dipole}). The other port of the VNA is connected to a single reference horn antenna that serves as a field probe. To position the SYNCA, a combination of translation and rotation stages are used to characterize the antenna's fields across the entire radiation pattern circumference. This measurement scheme is equivalent to measuring the fields generated by the SYNCA using a full circular array of probe antennas.

The antenna measurement space is surrounded by RF anti-reflective foam to isolate the measurements from the lab environment (see Figure \ref{fig:lab-meas-photo}) and remaining reflections are removed using the VNA's time-gating feature. The SYNCA is affixed to the stages by a custom RF transparent mount made of polystyrene foam. The coaxial cables deliver the antenna feed signals generated by the VNA to the SYNCA while still allowing unrestricted rotation. The horn antenna probe is nominally positioned in the plane formed by the antenna PCB ($\theta=90^\circ$ or $z=0$~mm) at a distance of 10~cm from the SYNCA, to match the expected position of the antenna array relative to the SYNCA in the antenna array test stand. The horn antenna can be manually raised or lowered to different relative vertical positions to characterize the radiation pattern at different polar angles.

\begin{figure}[h]
    \centering 
    \begin{subfigure}[b]{0.48\textwidth}
        \centering
        \includegraphics[width=\textwidth]{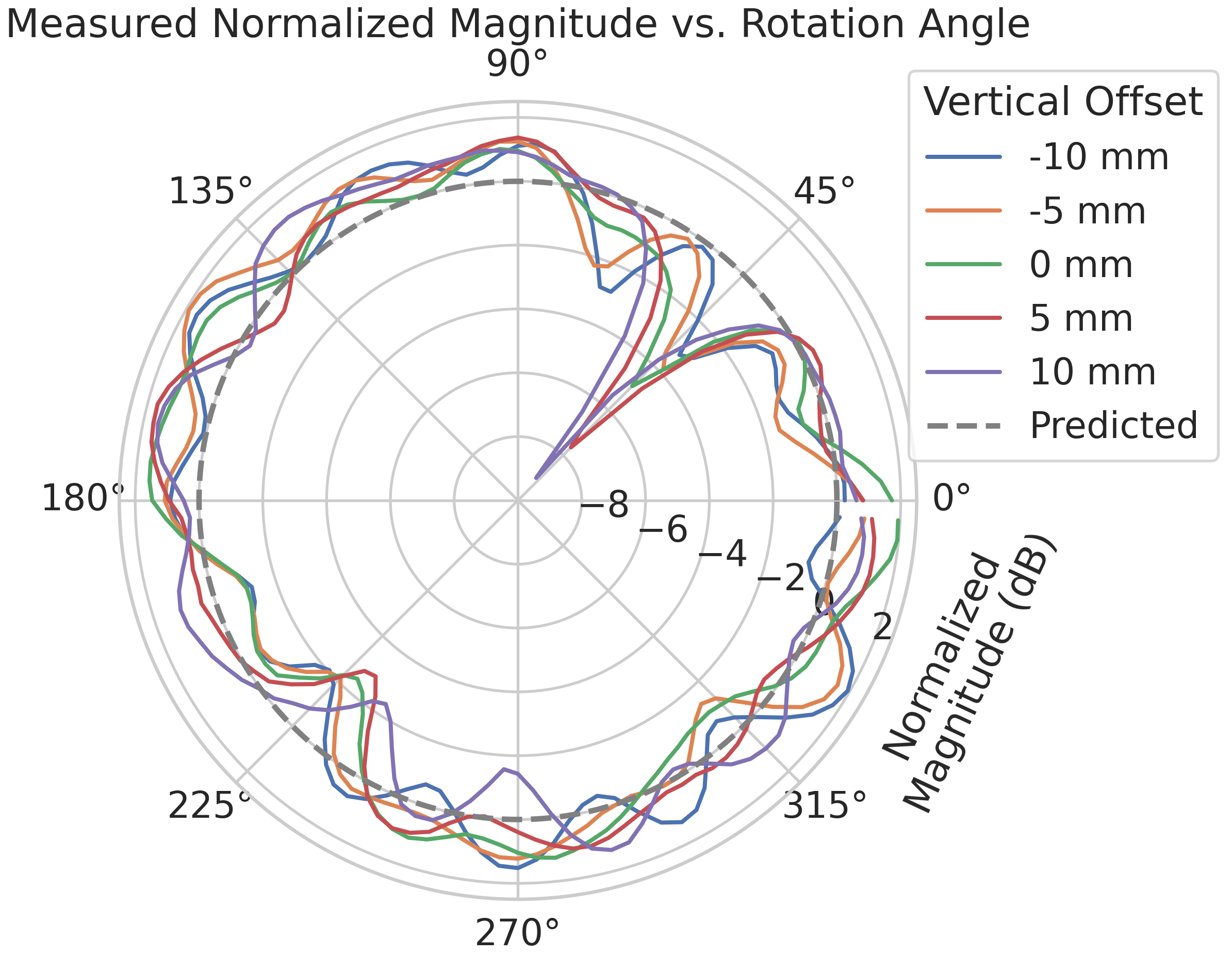}
        \caption{\label{fig:vna-meas-mag}}
    \end{subfigure}
    \hfill
    \begin{subfigure}[b]{0.48\textwidth}
        \centering
        \includegraphics[width=1.\textwidth]{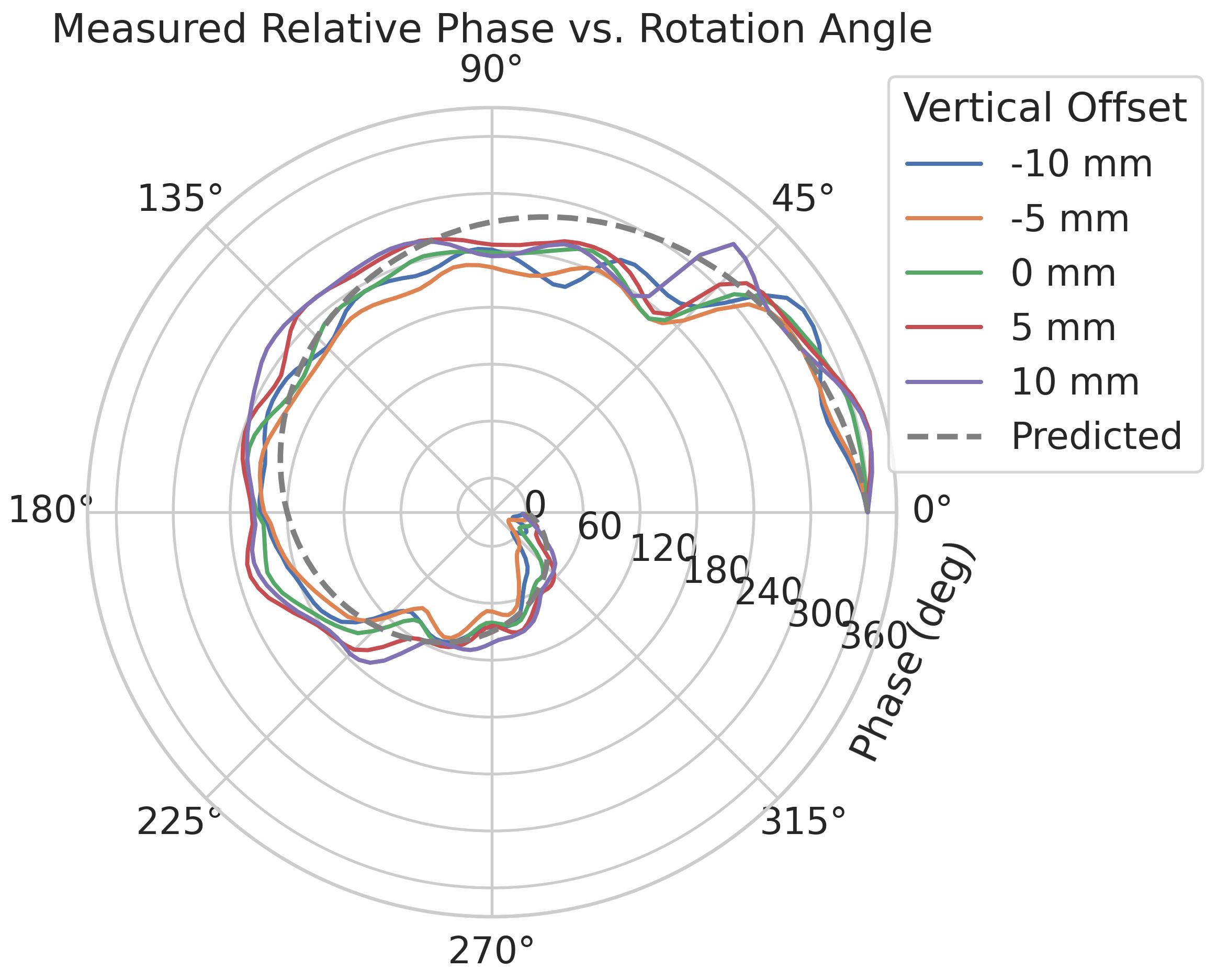}
        \caption{\label{fig:vna-meas-phase}}
    \end{subfigure}
    \hfill
    \caption{\label{fig:vna-meas} Linear interpolations of the measured electric field magnitude (a) and phase (b). The data was acquired using a VNA at 120 points spaced by 3 degrees from 0 to 357 degrees of azimuthal angle. The different color lines indicate the vertical offset of the horn antenna relative to the SYNCA PCB and the dashed line shows the expected shape from electron simulations. No significant difference in the antenna pattern is observed for the measured vertical offsets.}
    \qquad
\end{figure}

Several $360^\circ$ scans were performed with probe vertical offsets of -10.0~mm, -5.0~mm, 0.0~mm, 5.0~mm, and 10.0~mm relative to the antenna PCB plane. These probe offsets cover a 2~cm wide vertical region centered on the SYNCA PCB, approximately equal to $\pm6$ degrees of polar angle. The measurements show that the SYNCA is generating fields with nearly isotropic magnitude across the probed region. The standard deviation of the electric field magnitude measured around the antenna circumference is approximately 2.9~dB for a typical rotational scan.
The presence of a significant pattern null is noted near $45^\circ$ (see Figure \ref{fig:vna-meas}), which we attribute to small imperfections in the antenna PCB that could be introduced from the hand soldered terminations connecting the coax cables to the antenna. There is no significant difference in the radiation pattern when measured across the 2~cm vertical range. The measured relative phases closely follow the expectation for an electron, being linear with the measurement rotation angle and forming the expected spiral pattern. Other than the small phase imperfections there is a slight sinusoidal bias to the phase data, which we determined is the result of a small ($\lesssim 1$~mm) offset of the antenna's phase center from the rotation axis of the automated stages.

The characterization measurements confirm the simulated performance of the SYNCA. As expected the fields generated by the antenna are nearly isotropic in magnitude, $\phi$-polarized, and are linearly out of phase around the circumference of the antenna as predicted for cyclotron radiation in Section \ref{sec:pheno}. Small imperfections in the magnitude and phase of the antenna are expected, particularly at the antenna's high operating frequency of 26~GHz where small geometric changes can have significant impacts on electrical properties. However, calibration through careful characterization measurements can be used to remove the majority of these pattern imperfections, including the relatively large pattern null near $45^\circ$, which will allow for the usage of the SYNCA as a test source for free-space CRES experiments utilizing antenna arrays. In the next section we use the VNA measurements obtained here as a calibration for signal reconstruction using digital beamforming.

\section{Signal Reconstruction with the SYNCA}

\begin{figure}[b]
    \centering
    \begin{subfigure}[b]{0.45\textwidth}
        \centering
        \includegraphics[width=.85\textwidth]{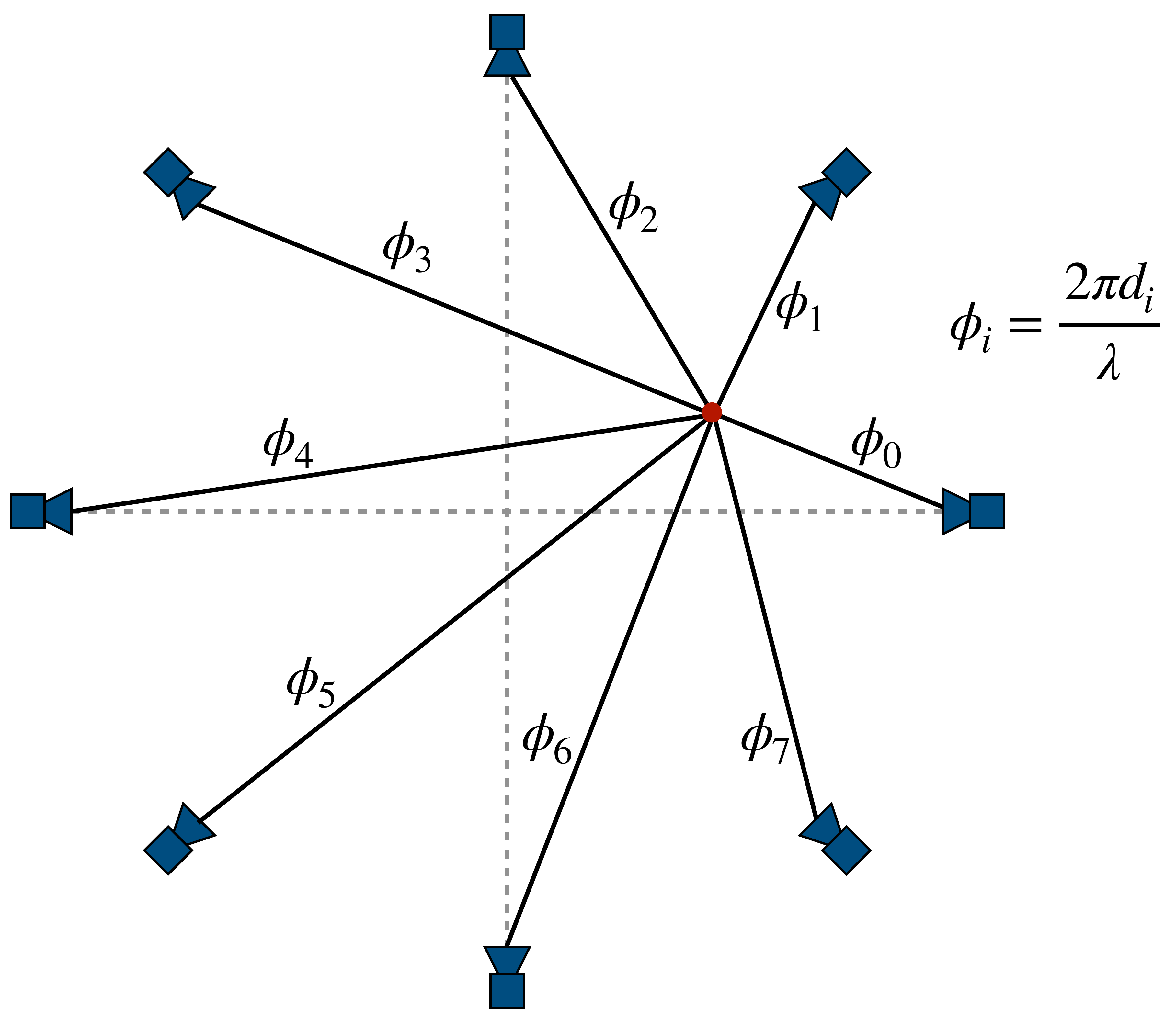}
        \caption{\label{fig:beamforming-concept}}
    \end{subfigure}
    \hfill
    \begin{subfigure}[b]{0.45\textwidth}
        \centering
        \includegraphics[width=\textwidth]{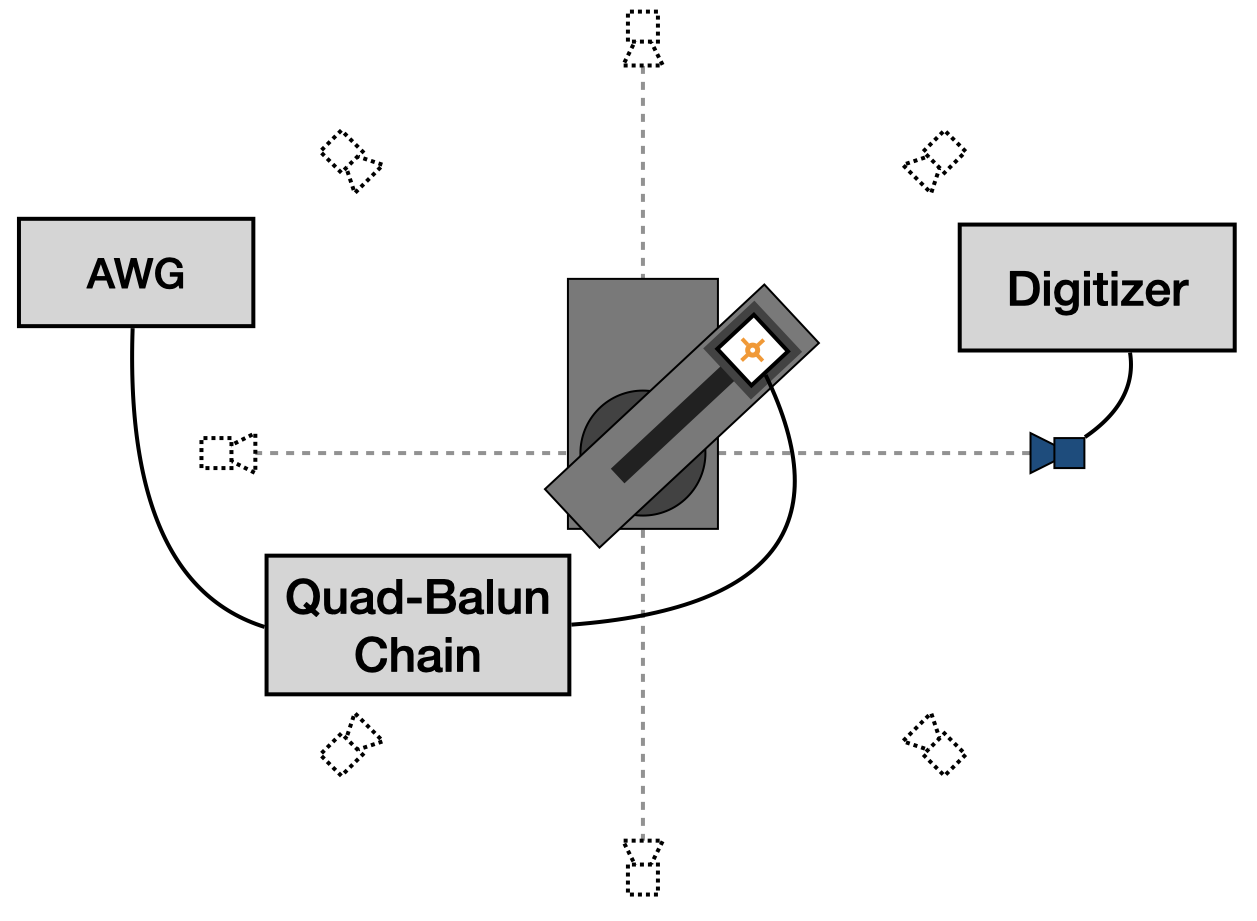}
        \caption{\label{fig:beamforming-measurement}}
    \end{subfigure}
    \hfill
    \caption{\label{fig:beamforming} (a) A depiction of the relative phase differences for signals received by a circular antenna array from an isotropic source. The phases correspond to a unique spatial position. (b) A schematic of the setup used to perform digital beamforming.}
    \qquad
\end{figure}

Digital beamforming is a standard technique for signal reconstruction using a phased array \cite{wirth2001radar}. The SYNCA, since it exhibits the same cyclotron phases as a trapped electron, can be used to perform simulated CRES digital beamforming reconstruction experiments on the bench-top without the need for the magnet, cryogenics, and vacuum systems required by a full CRES experiment. The fields received by the individual elements of the antenna array will have phases dependent on the spatial position of the source relative to the antennas. Therefore, a simple summation of the received signals will fail to reconstruct the signal due to destructive interference between the individual channels in the array. However, applying a phase shift associated with the source's spatial position removes phase differences and results in a constructive summation of the channel signals (see Figure \ref{fig:beamforming}). We can summarize the digital beamforming operation succinctly using the following equation
\begin{equation}
    y[t_n]=\sum_{m=0}^{N-1}x_m[t_n]A_m e^{i\phi_m},
\end{equation}
where $y[t_n]$ represents the summed array signal at time $t_n$, $x_m[t_n]$ is the signal received by channel $m$ at time $t_n$, $\phi_m$ is the phase shift applied to the signal received at channel $m$, and $A_m$ is an amplitude weighting factor that accounts for the different signal power received by individual channels. By changing the digital beamforming phases, the point of constructive interference can be scanned across the sensitive region of the array to search for the location of a radiating source, which is identified as the point of maximum summed signal power above a specified threshold. The digital beamforming phases consist of two components,
\begin{equation}
    \phi_m=2\pi d_m/\lambda+\theta_m,
    \label{eq:beamform-phase}
\end{equation}
where $d_m$ is the distance from the $m$-th array element to the source, and $\theta_m$ is the relative angle between the source position and the $m$-th antenna. The first component is the standard digital beamforming phase that corresponds to the spatial position of the source, and the second component is the cyclotron phase that corresponds to the relative azimuthal phase offset. 

\begin{figure}[h]
    \centering
    \includegraphics[width=0.8\textwidth]{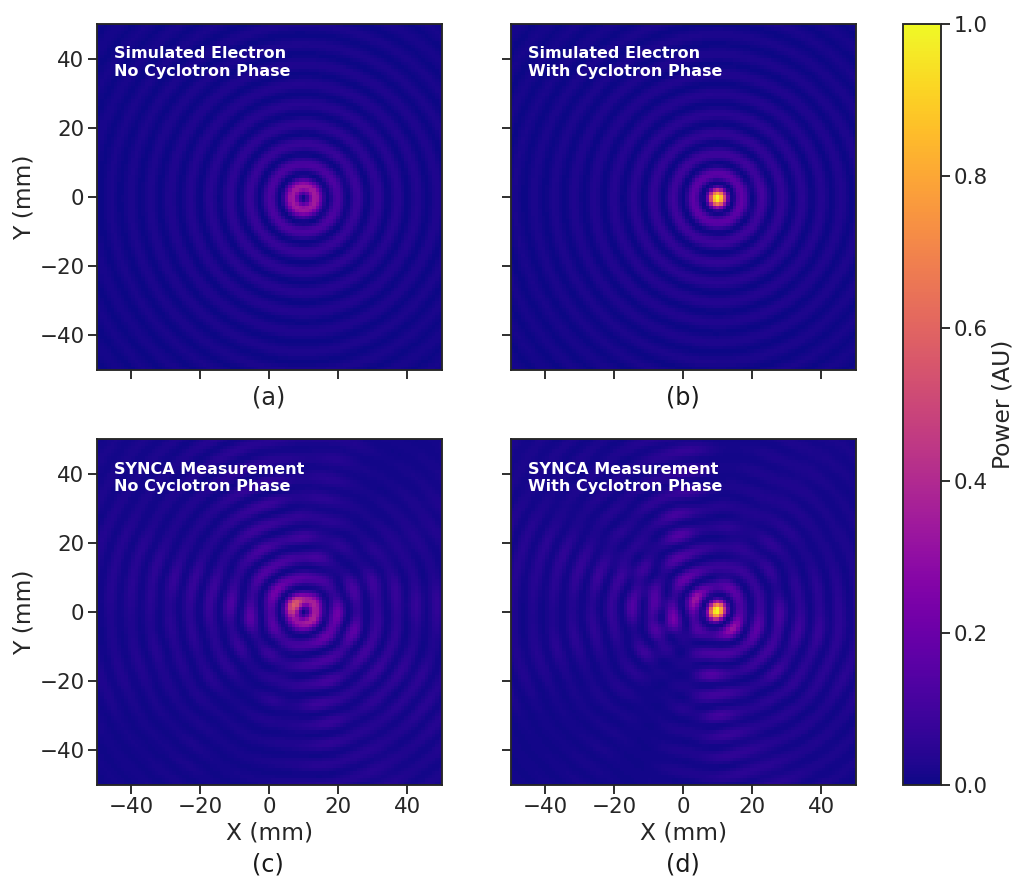}
    \caption{\label{fig:bf} Digital beamforming maps generated using a simulated 60 channel array and electron simulated using the Locust package. (a) and (b) show the beamforming maps for simulated electrons without the cyclotron spiral phases and with the cyclotron spiral phases respectively. (c) and (d) show the beamforming maps produced from SYNCA measurements. We observe good agreement between simulated electrons and the SYNCA measurements.}
    \qquad
\end{figure}

With a small modification to the hardware used to characterize the SYNCA (see Figure \ref{fig:vna-meas-schematic}), we can perform a digital beamforming reconstruction of a synthetic CRES event. By replacing the VNA with an arbitrary waveform generator (AWG), the SYNCA can be used to generate cyclotron radiation with an arbitrary signal structure, which can then be detected by digitizing the signals received by the horn antenna. Rotational symmetry allows us to use the rotational stage of the positioning system to rotate the SYNCA to recreate the signals that would have been received by a complete circular array of antennas. 

Using this setup, signals from a 60 channel circular array of equally spaced horn antennas were generated with the SYNCA positioned 10~mm off the central array axis, reconstructed using digital beamforming, and compared to Locust simulation (see Figure \ref{fig:bf}). When the cyclotron spiral phases are not used, which is equivalent to setting $\theta_m$ in Equation \ref{eq:beamform-phase} to zero, the SYNCA's position is reconstructed as a relatively faint ring as predicted by simulation. However, when the appropriate cyclotron phases are used during the beamforming procedure, both the simulated electron and the SYNCA appear as a single peak of high relative power corresponding to the source position. Therefore, we observe good agreement between the simulated and SYNCA reconstructions. While it may seem that for the case with no cyclotron phase corrections the ring reconstructs the position of the electron as effectively as beamforming with the cyclotron phase corrections, it is important to note that the simulations and measurements were generated without a realistic level of thermal noise. The larger maxima region and lower signal power, which occurs without the cyclotron phase corrections, significantly reduce the probability of detecting an electron in a realistic noise background.
\begin{figure}[htbp]
\centering 
    \begin{subfigure}[b]{0.48\textwidth}
        \centering
        \includegraphics[width=1\textwidth]{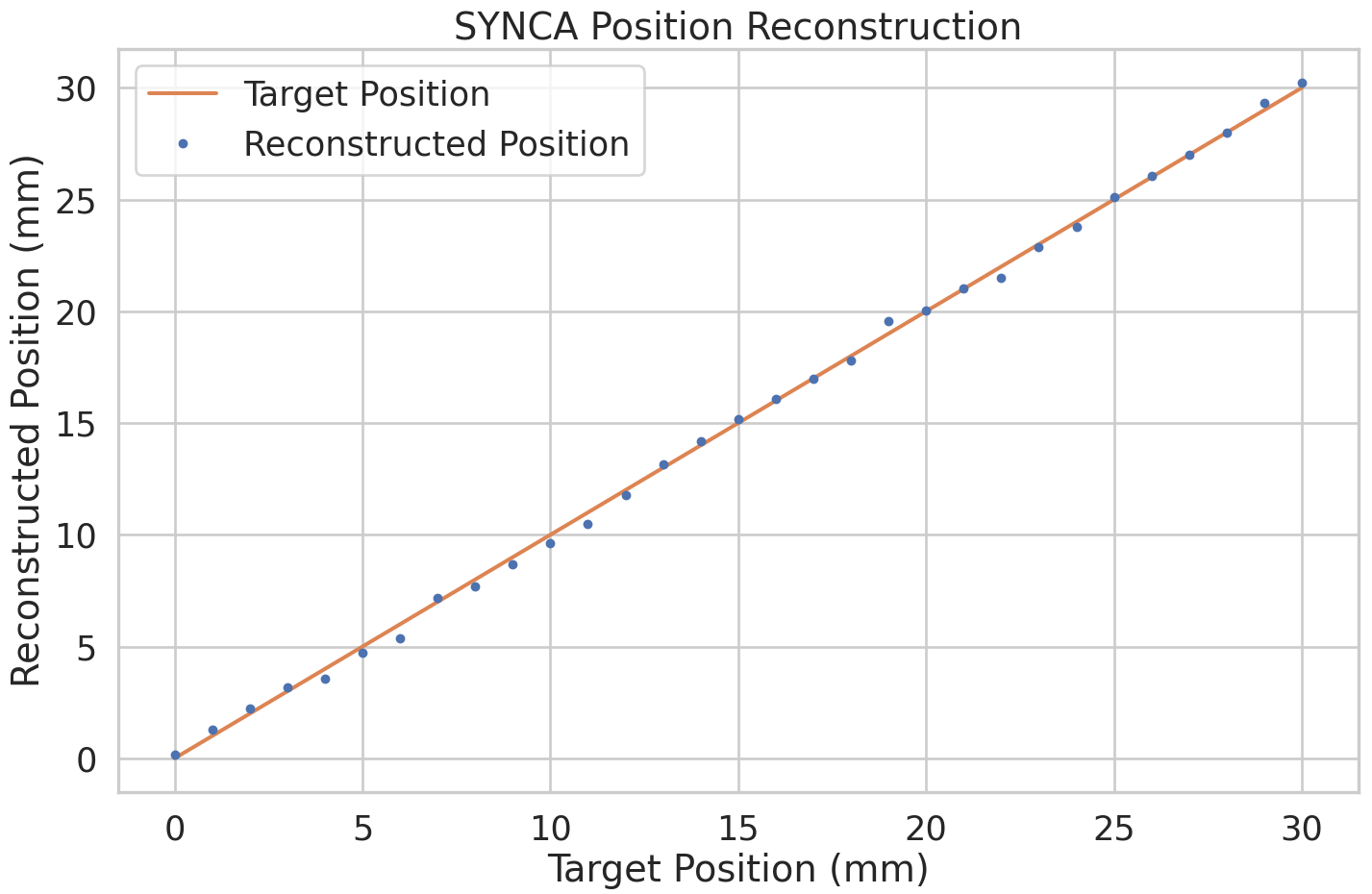}
        \caption{\label{fig:synca-sweep-recon}}
    \end{subfigure}
    \hfill
    \begin{subfigure}[b]{0.48\textwidth}
        \centering
        \includegraphics[width=\textwidth]{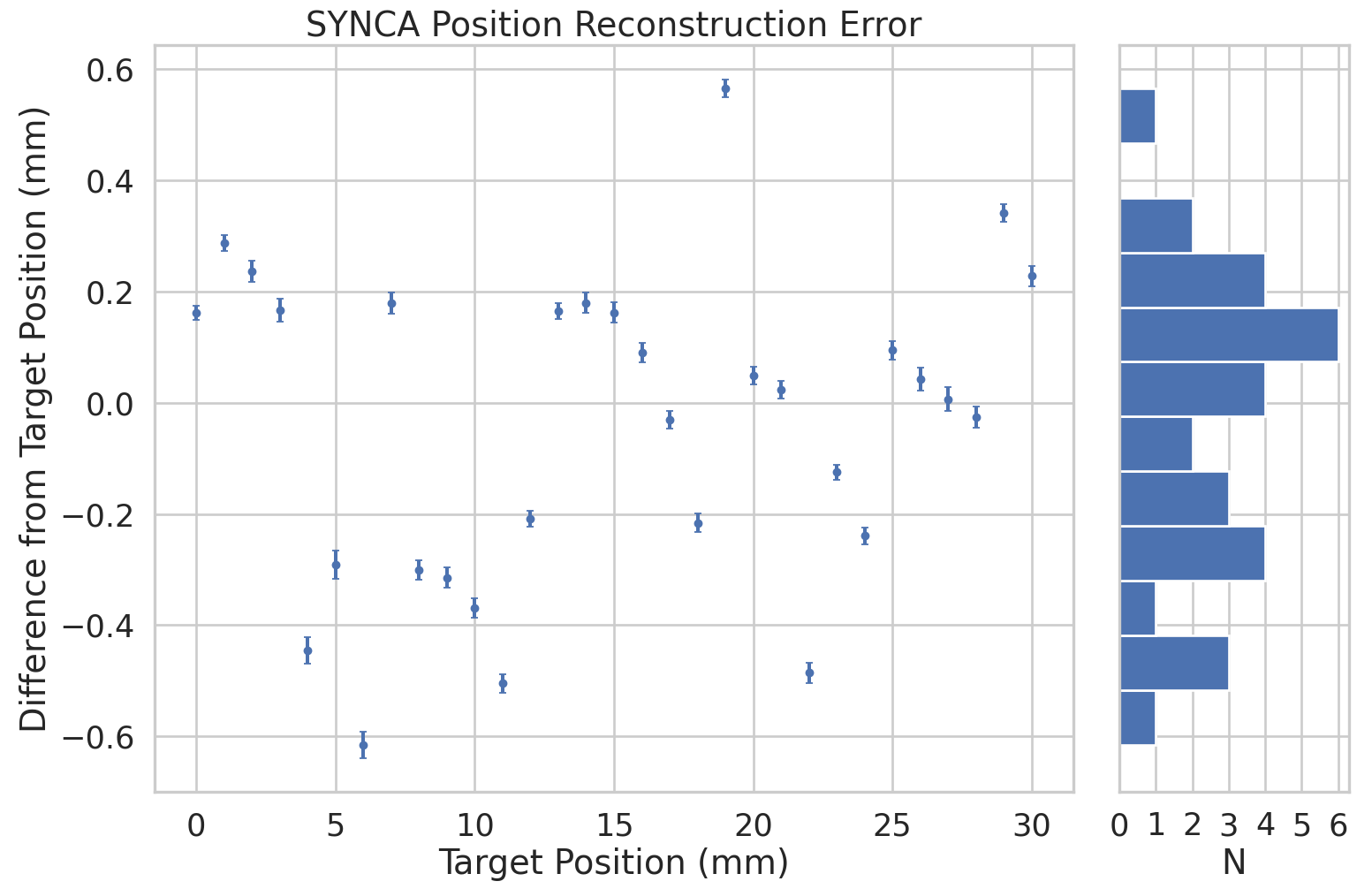}
        \caption{\label{fig:synca-sweep-error}}
    \end{subfigure}
    \hfill
\caption{\label{fig:synca-sweep} A plot of the SYNCA's reconstructed position using the synthesized horn-antenna array and digital beamforming. (a) Shows the reconstructed position of the SYNCA compared with the target position indicated by the positioning system readout. (b) Shows the reconstruction error, which is the difference between the target and reconstructed positions. The error bars in (b) are the uncertainty in the mean position of the 2D Gaussian used to fit the digital beamforming reconstruction peak obtained from the fit covariance matrix. The mean fit position uncertainty of 0.02~mm is an order of magnitude smaller than the typical reconstruction error of 0.3~mm obtained by calculating the standard deviation of the difference between the reconstructed and target position.}
\end{figure}

To bound the beamforming capabilities of the synthetic array of horn antennas, we performed a series of beamforming reconstructions where the SYNCA was progressively moved off the central axis of the array (see Figure \ref{fig:synca-sweep}). To extract an estimate of the position of the SYNCA using the digital beamforming image we apply a 2-dimensional (2D) Gaussian fit to the image data and extract the estimated centroid value. We find that the synthetic horn antenna array reconstructs the position of the SYNCA with a $1\sigma$-error of 0.3~mm with no apparent trend across the 30~mm measurement range. This reconstruction error is an order of magnitude larger than mean fit position uncertainty of 0.02~mm indicating that systematic effects related to the SYNCA positioning system could be contributing additional uncertainty to the measurements. Note that the current mean reconstruction error of 0.3~mm is a factor of 20 smaller than the full width at half maximum of the digital beamforming peak (6~mm), which could be interpreted as a naive estimate of the position reconstruction performance of this technique. Because these measurements are intended as a proof-of-principle demonstration, we do not investigate potential sources of systematic errors further; however, we expect that a similar and more thorough investigation will be performed using the Project 8 antenna array test stand, where typical reconstruction errors can be used to estimate the energy resolution limits of antenna array designs.

\section{Conclusions}

In this paper we have introduced the SYNCA, which is a novel antenna design that emits radiation that mimics the unique properties of the cyclotron radiation generated by charged particles moving in a magnetic field. The characterization measurements of the SYNCA validated the simulated performance of the PCB crossed-dipole antenna design. Additionally, the SYNCA was used to estimate the position reconstruction capabilities of a synthesized array of horn antennas and experimentally reproduced the simulated digital beamforming reconstruction of electrons. 

While the SYNCA performs well, there exist discrepancies in the phase and magnitude of the radiation pattern compared to the simulated SYNCA design that are related to the small geometric differences in the soldered connections. Future design iterations that replace the soldered connections with a fully surface mount design could improve the radiation pattern at the cost of some complexity and expense. Furthermore, improving the design of the antenna PCB and mounting system would allow the antenna to be inserted into a cryogenic and vacuum environment where in-situ antenna measurement calibrations could be performed. 

The discrepancies in the radiation pattern and phases exhibited by the as-built SYNCA should not greatly impact its performance as a calibration probe. Both magnitude and phase variations can be accounted by applying the SYNCA characterization measurements as a calibration to the data collected by the antenna array test stand. The separate calibration of the SYNCA radiation does not impact the primary goals for the antenna array test stand which are array calibration and signal reconstruction algorithm performance characterization, because it can be performed with standard reference horn antennas with well understood characteristics.

The SYNCA antenna technology advances the CRES technique by providing a mechanism to characterize free-space antenna arrays for CRES measurements without the need for a magnet and cryogenics system, which would be required for calibration using electron sources. Both the Project 8 collaboration as well as future collaborations which are developing antenna array based CRES experiments can make use of SYNCA antennas as an important component of their calibration and commissioning phases.

\appendix
\section{Dominant Electric Field Polarization}
\label{sec:app-1}
Consider an electron following a circular cyclotron orbit in a uniform magnetic field whose guiding center is positioned at the origin of the coordinate system. The equation of motion can be expressed as 
\begin{equation}
    \vec r_s= (r_c\cos{\omega_ct_{r}})\hat x+(r_c\sin{\omega_ct_{r}})\hat y.
\end{equation}
For single antenna located along the y-axis at position $\vec{r}=r_{a}\hat{y}$ we are interested in the incident electric fields from the electron. The electric field is given by Equation \ref{eq:lw-e}, which we evaluate in the regime where $r_{a}\gg r_c$. This limit can justified by comparing the radius of the cyclotron orbit for an electron with the tritium beta-spectrum endpoint energy of 18.6~keV in a 1~T magnetic field to the typical ($r_{a}\simeq100$~mm) radial position of the receiving antenna. We find that the cyclotron orbit has a radius of 0.46~mm which is approximately a factor of 200 smaller than the typical antenna radial position. In this regime we can make the approximation $\vec{R}\simeq r_{a}\hat{y}$ and the expression for the electric field at the antenna's position becomes
\begin{multline}
    \label{eq:lw-xy-comp}
    \vec{E}=\frac{e}{\gamma^2r_{a}^2}\frac{\hat{x}(\frac{r_c\omega_c}{c}\sin{\omega_ct_{r}})+\hat{y}(1-\frac{r_c\omega_c}{c}\cos{\omega_ct_{r}})}{(1-\frac{r_c\omega_c}{c}\cos{\omega_ct_{r}})^3}
    - \frac{e}{cr_{a}}\frac{\hat{x}(\frac{r_c^2\omega_c^3}{c^2}-\frac{r_c\omega_c^2}{c}\cos{\omega_ct_{r}})}{(1-\frac{r_c\omega_c}{c}\cos{\omega_ct_{r}})^3}.
\end{multline}
Since the receiving antenna is part of a circular array of antennas, it is useful to rewrite Equation \ref{eq:lw-xy-comp} in terms of the azimuthal ($\hat{\phi}$) and radial ($\hat{r}$) polarizations. Making use of the fact that for an antenna located at $R=r_{a}\hat{y}$ that $\hat{\phi}=-\hat{x}$ and $\hat{r}=\hat{y}$ we find
\begin{align}
    \vec{E}&=\hat{\phi}E_\phi+\hat{r}E_r\\
    \label{eq:e-field-azimuthal-comp}
    E_\phi&=\frac{e}{(1-\frac{r_c\omega_c}{c}\cos{\omega_ct_{r}})^3}\left[-\frac{\frac{r_c\omega_c}{c}\sin{\omega_ct_{r}}}{\gamma^2r_{a}^2}+\frac{\omega_c\left(\frac{r_c^2\omega_c^2}{c^2}-\frac{r_c\omega_c}{c}\cos{\omega_ct_{r}}\right)}{cr_{a}}\right]\\
    E_r&=\frac{e(1-\frac{r_c\omega_c}{c}\sin{\omega_ct_{r}})}{\gamma^2r_{a}^2(1-\frac{r_c\omega_c}{c}\cos{\omega_ct_{r}})^3}.
\end{align}

For the purposes of designing a synthetic cyclotron radiation antenna we are interested in the dominant electric field polarization emitted by the electron. The antenna is being designed to mimic the cyclotron radiation produced by electrons with kinetic energies of approximately 18.6~keV in a 1~T magnetic field. Since the relativistic beta factor for an electron with this kinetic energy is  $|\vec{\beta}|\simeq\frac{1}{4}$, the approximations $\gamma\simeq1$ and $\frac{r_c\omega_c}{c}\simeq\frac{1}{4}$ are justified. Inserting these expressions into the equations for the electric field components above simplifies the comparison of the magnitudes of the two components. Additionally, we compare the time-averaged magnitudes to evaluate the root mean squared electric field ratio. The time-averaged ratio of the radial and azimuthally polarizied electric fields with the above simplifications is given by
\begin{equation}
    \label{eq:e-field-ratio}
    \frac{\left<|E_r|\right>}{\left<|E_\phi|\right>}=\frac{8-\sqrt{2}}{\left|1-\frac{r_{a}}{r_c}\frac{1-2\sqrt{2}}{8}\right|}\simeq\frac{r_c}{r_{a}}\frac{8(8-\sqrt{2})}{2\sqrt{2}-1}=0.13,
\end{equation}
where we have made use of the fact that for these magnetic fields and kinetic energies the cyclotron radius is much smaller than the radius of the antenna array.

From Equation \ref{eq:e-field-ratio} we see that the time-averaged azimuthal polarization is larger than the radial polarization by about a factor of 8, which makes it the dominant contribution to the electric fields at the position of the antenna. We must also consider the directivity of the receiving antenna which can have a gain that is disproportionately large for a specific polarization component. Because the $E_\phi$ component is dominant, the receiving antenna array is designed with an azimuthal polarization, which negates the voltages induced in the antenna from the radially polarized fields. Therefore, we conclude that for the purpose of designing the SYNCA antenna it is sufficient to specify that the emitted electric fields have a large azimuthal polarization.

\section{Simplification of the Expression for the Azimuthally Polarized Electric Fields}
\label{sec:app-2}
\begin{figure}[htbp]
    \centering
    \includegraphics[width=\textwidth]{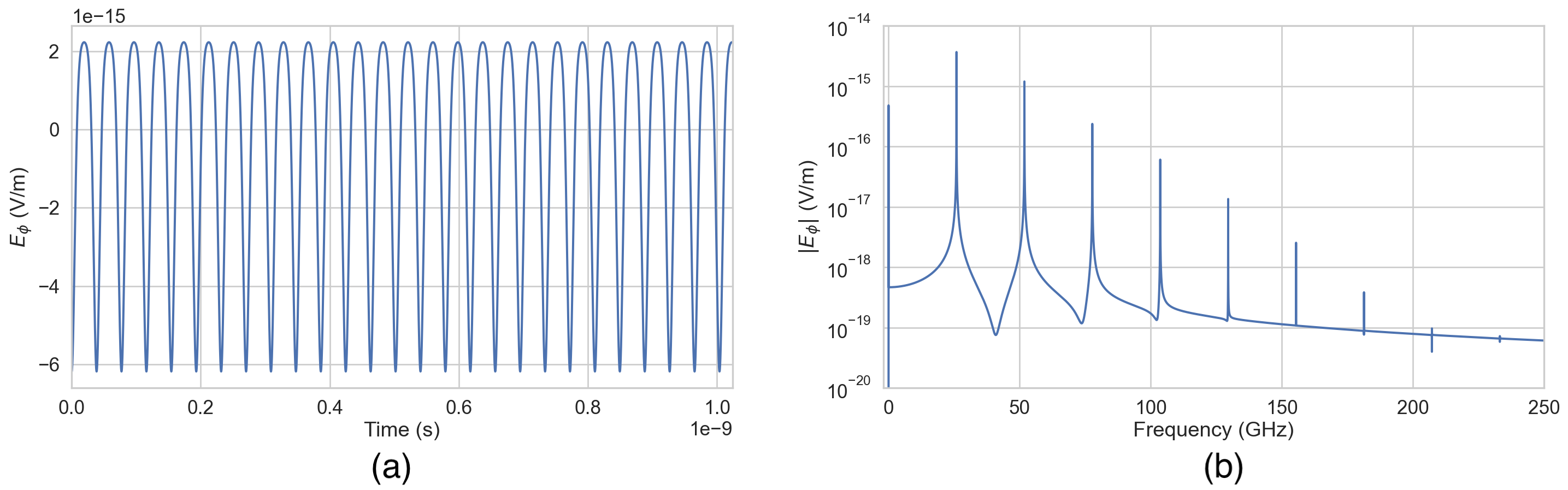}
    \caption{A plot of the numeric solution to Equation \ref{eq:e-field-azimutal-comp-simple}. The time-domain representation of the signal (a) is composed of a zero frequency term and a series of harmonics separated by the main cyclotron frequency as shown in the plot of the frequency spectrum (b). We can see that the relative amplitude of the harmonics beyond $k=7$ are smaller than the main carrier by a factor of about $10^{-5}$ and are completely negligible.}
    \label{fig:lw-azimuth-time-spectrum}
\end{figure}

Returning to Equation \ref{eq:e-field-azimuthal-comp} we can rearrange the terms into a more compact form
\begin{equation}
    \label{eq:e-field-azimutal-comp-simple}
    E_\phi = \frac{e\frac{r_c\omega_c}{c}}{4r_{a}r_c}\left[\frac{\frac{r_c\omega_c}{c}-\cos{\omega_ct_{r}}-\frac{4r_c}{ r_{a}}\sin{\omega_ct_{r}}}{(1-\frac{r_c\omega_c}{c}\cos{\omega_ct_{r}})^3}\right],
\end{equation}
where we have used the relation $\frac{c}{\omega_c}\simeq4\gamma r_c$, which is true for electrons with these kinetic energies, to replace the extra factor of $\omega_c$ and replaced the relativistic gamma factor with $1$. Recall, that this equation is valid for an electron moving in a circular trajectory with radius $r_c$ whose electric fields are being received by an antenna located in the plane of the cyclotron orbit at a distance $r_a$ from the electron's guiding center. This rather complicated expression can be simplified using Fourier analysis. Assuming a background magnetic field of 1~T and a kinetic energy of 18.6~keV we calculate numerically the electric field using Equation \ref{eq:e-field-azimutal-comp-simple} and apply a discrete Fourier Transform to visualize the frequency spectrum (see Figure \ref{fig:lw-azimuth-time-spectrum}).

We observe that the azimuthally polarized electric field is periodic with a base cyclotron frequency of 25.898~GHz corresponding to the highest power frequency component in Figure \ref{fig:lw-azimuth-time-spectrum}. The frequency spectrum reveals that the signal is composed of a constant term with zero frequency and a series of harmonics separated by 25.898~GHz. Therefore, we can represent the azimuthal electric fields from the electron as a linear combination of pure sinusoids with frequencies given by $\omega_k=k\omega_c$ ($k\in0,1,2...$) and amplitudes extracted from the Fourier representation. The exponential decay of the amplitudes implies that it will usually be sufficient to truncate the sum after the 4th harmonic whose amplitude is approximately a factor of 100 less than the 1st order frequency term. Using this representation we can transform the equation for the azimuthally polarized electric fields in Equation \ref{eq:e-field-azimutal-comp-simple} into 
\begin{equation}
    \label{eq:e-field-azimutal-comp-final}
    E_\phi = \frac{e\frac{r_c\omega_c}{c}}{4r_{a}r_c}\sum_{k=0}^7{A_k e^{i\omega_k t_{r}}},
\end{equation}
where we have truncated the sum over harmonics at the 7th order for completeness. The amplitudes $A_k$ are dimensionless complex numbers, which encode the relative powers of the harmonics as well as the starting overall phase of the cyclotron radiation.

\acknowledgments

A special thanks to Field Theory Consulting, which helped to design the PCB crossed-dipole antenna and fabricated the SYNCA's. 

This material is based upon work supported by the following sources: the U.S. Department of Energy Office of Science, Office of Nuclear Physics, under Award No.~DE-SC0020433 to Case Western Reserve University (CWRU), under Award No.~DE-SC0011091 to the Massachusetts Institute of Technology (MIT), under Field Work Proposal Number 73006 at the Pacific Northwest National Laboratory (PNNL), a multiprogram national laboratory operated by Battelle for the U.S. Department of Energy under Contract No.~DE-AC05-76RL01830, under Early Career Award No.~DE-SC0019088 to Pennsylvania State University, under Award No.~DE-FG02-97ER41020 to the University of Washington, and under Award No.~DE-SC0012654 to Yale University; the National Science Foundation under Award No.~PHY-2209530 to Indiana University, and under Award No.~PHY-2110569 to MIT; This work has been supported by the Cluster of Excellence "Precision Physics, Fundamental Interactions, and Structure of Matter" (PRISMA+ EXC 2118/1) funded by the German Research Foundation (DFG) within the German Excellence Strategy (Project ID 39083149); the Karlsruhe Institute of Technology (KIT) Center Elementary Particle and Astroparticle Physics (KCETA); Laboratory Directed Research and Development (LDRD) 18-ERD-028 and 20-LW-056 at Lawrence Livermore National Laboratory (LLNL), prepared by LLNL under Contract DE-AC52-07NA27344, LLNL-JRNL-840701; the LDRD Program at PNNL; Indiana University; and Yale University.  A portion of the research was performed using the HPC cluster at the Yale Center for Research Computing. Computations for this research were performed on the Pennsylvania State University’s Institute for Computational and Data Sciences’ Roar supercomputer.

\bibliographystyle{JHEP}
\bibliography{refs}
\end{document}